\documentclass[journal]{IEEEtran}
\normalsize
\usepackage{cite}
\usepackage[cmex10]{amsmath}
\usepackage{amssymb}
\usepackage{amsthm}
\usepackage{mathrsfs}
\usepackage{bm}
\usepackage[mathscr]{eucal}
\usepackage{amssymb,amsmath,amsthm,amsfonts,latexsym}
\usepackage{amsmath,graphicx,bm,xcolor,url}
\usepackage{graphicx}
\usepackage{latexsym}
\usepackage{CJK}
\usepackage{indentfirst}
\usepackage{psfrag}
\usepackage{setspace}
\usepackage{algorithmic}
\usepackage{algorithmic, cite}
\usepackage{algorithm}
\usepackage{array}
\usepackage{mdwmath}
\usepackage{mdwtab}
\usepackage{eqparbox}
\usepackage{url}
\usepackage{epstopdf}
\usepackage{epsfig,epsf,psfrag}
\usepackage{fixltx2e}
\usepackage{verbatim}
\usepackage{textcomp}
\usepackage{psfrag} 

\usepackage{subfigure} 
\usepackage{caption}

\theoremstyle{plain}
\newtheorem{mythe}{Theorem}
\theoremstyle{remark}

\theoremstyle{plain}

\theoremstyle{remark}

\theoremstyle{plain}

\theoremstyle{remark}

\theoremstyle{remark}

\theoremstyle{remark}

\theoremstyle{remark}

\theoremstyle{remark}

\theoremstyle{remark}

\catcode`~=11 \def\UrlSpecials{\do\~{\kern -.15em\lower .7ex\hbox{~}\kern .04em}} \catcode`~=13

\allowdisplaybreaks[4]

\newcommand{\calA}{\mathcal{A}}

\newcommand{\calC}{\mathcal{C}}
\newcommand{\calD}{\mathcal{D}}

\newcommand{\calN}{\mathcal{N}}
\newcommand{\calO}{\mathcal{O}}

\newcommand{\calQ}{\mathcal{Q}}

\newcommand{\ba}{\mathbf{a}}
\newcommand{\bA}{\mathbf{A}}
\newcommand{\bb}{\mathbf{b}}

\newcommand{\bc}{\mathbf{c}}
\newcommand{\bC}{\mathbf{C}}

\newcommand{\boldf}{\mathbf{f}}
\newcommand{\bF}{\mathbf{F}}
\newcommand{\bg}{\mathbf{g}}
\newcommand{\bG}{\mathbf{G}}
\newcommand{\bh}{\mathbf{h}}
\newcommand{\bH}{\mathbf{H}}

\newcommand{\bI}{\mathbf{I}}

\newcommand{\bs}{\mathbf{s}}
\newcommand{\bS}{\mathbf{S}}

\newcommand{\bU}{\mathbf{U}}

\newcommand{\bW}{\mathbf{W}}
\newcommand{\bx}{\mathbf{x}}

\newcommand{\bY}{\mathbf{Y}}

\newcommand{\rmb}{\mathrm{b}}

\newcommand{\rmd}{\mathrm{d}}

\newcommand{\rmF}{\mathrm{F}}

\newcommand{\rmH}{\mathrm{H}}

\newcommand{\rmn}{\mathrm{n}}

\newcommand{\rmp}{\mathrm{p}}

\newcommand{\rmT}{\mathrm{T}}

\newcommand{\bbC}{\mathbb{C}}

\newcommand{\bbE}{\mathbb{E}}

\DeclareMathAlphabet{\mathbsf}{OT1}{cmss}{bx}{n}
\DeclareMathAlphabet{\mathssf}{OT1}{cmss}{m}{sl}

\DeclareSymbolFont{bsfletters}{OT1}{cmss}{bx}{n}
\DeclareSymbolFont{ssfletters}{OT1}{cmss}{m}{n}
\DeclareMathSymbol{\bsfGamma}{0}{bsfletters}{'000}
\DeclareMathSymbol{\ssfGamma}{0}{ssfletters}{'000}
\DeclareMathSymbol{\bsfDelta}{0}{bsfletters}{'001}
\DeclareMathSymbol{\ssfDelta}{0}{ssfletters}{'001}
\DeclareMathSymbol{\bsfTheta}{0}{bsfletters}{'002}
\DeclareMathSymbol{\ssfTheta}{0}{ssfletters}{'002}
\DeclareMathSymbol{\bsfLambda}{0}{bsfletters}{'003}
\DeclareMathSymbol{\ssfLambda}{0}{ssfletters}{'003}
\DeclareMathSymbol{\bsfXi}{0}{bsfletters}{'004}
\DeclareMathSymbol{\ssfXi}{0}{ssfletters}{'004}
\DeclareMathSymbol{\bsfPi}{0}{bsfletters}{'005}
\DeclareMathSymbol{\ssfPi}{0}{ssfletters}{'005}
\DeclareMathSymbol{\bsfSigma}{0}{bsfletters}{'006}
\DeclareMathSymbol{\ssfSigma}{0}{ssfletters}{'006}
\DeclareMathSymbol{\bsfUpsilon}{0}{bsfletters}{'007}
\DeclareMathSymbol{\ssfUpsilon}{0}{ssfletters}{'007}
\DeclareMathSymbol{\bsfPhi}{0}{bsfletters}{'010}
\DeclareMathSymbol{\ssfPhi}{0}{ssfletters}{'010}
\DeclareMathSymbol{\bsfPsi}{0}{bsfletters}{'011}
\DeclareMathSymbol{\ssfPsi}{0}{ssfletters}{'011}
\DeclareMathSymbol{\bsfOmega}{0}{bsfletters}{'012}
\DeclareMathSymbol{\ssfOmega}{0}{ssfletters}{'012}

\newcommand{\hatc}{\widehat{c}}

\newcommand{\hatbc}{\widehat{\bc}}

\newcommand{\tilH}{\widetilde{H}}

\newcommand{\hatbH}{\widehat{\bH}}
\newcommand{\tilbh}{\widetilde{\bh}}
\newcommand{\tilbH}{\widetilde{\bH}}

\newcommand{\hatS}{\widehat{S}}

\newcommand{\hatbS}{\widehat{\bS}}

\newcommand{\bepsilon}{\bm{\epsilon}}

\newcommand{\bSigma	}{\bm{\Sigma}}

\def\norm#1{\left\| #1 \right\|}
\def\norm2#1{\left\| #1 \right\|_2}
\def\norm22#1{\left\| #1 \right\|_2^2}

\DeclareMathOperator{\diag}{diag}

\newcommand{\qednew}{\nobreak \ifvmode \relax \else
      \ifdim\lastskip<1.5em \hskip-\lastskip
      \hskip1.5em plus0em minus0.5em \fi \nobreak
      \vrule height0.75em width0.5em depth0.25em\fi}

\usepackage{float}
\usepackage{cite}
\usepackage{times,amsmath,color,amssymb,graphicx,epsfig,multirow,float,algorithm,algorithmic}
\usepackage{caption}
\usepackage{subfigure}
\usepackage[utf8]{inputenc}

\newtheorem{remark}{Remark}
\usepackage[hidelinks]{hyperref}

\begin{document}
\captionsetup{font={small}}
\title{Pilot Design and Signal Detection for Symbiotic Radio over OFDM Carriers}

\author{Hao Chen, Qianqian Zhang, Ruizhe Long, Yiyang Pei, \IEEEmembership{Senior Member,~IEEE}, and Ying-Chang Liang, \IEEEmembership{Fellow,~IEEE}
\thanks{Part of this work was presented in IEEE GLOBECOM'22 \cite{conference}. This work was supported in part by the National Key Research and Development Program of China under Grant 2018YFB1801105; in part by the Key Areas of Research and Development Program of Guangdong Province, China, under Grant 2018B010114001; in part by the Fundamental Research Funds for the Central Universities under Grant ZYGX2019Z022; and in part by the Program of Introducing Talents of Discipline to Universities under Grant B20064. \emph{(Corresponding author: Ying-Chang Liang.)}}
\thanks{H.~Chen is with the National Key Laboratory of Wireless Communications, University of Electronic Science and Technology of China, Chengdu 611731, P. R. China, and also with the Yangtze Delta Region Institute (Huzhou), University of Electronic Science and Technology of China, Huzhou 313001, P. R. China (e-mail: {hhhaochen@std.uestc.edu.cn}).}
\thanks{Q.~Zhang and R.~Long are with the National Key Laboratory of Wireless Communications, University of Electronic Science and Technology of China, Chengdu 611731, P. R. China (e-mail: {qqzhang\_kite@163.com;~ruizhelong@gmail.com}).}
\thanks{Y.-C.~Liang is with the Center for Intelligent Networking and Communications (CINC), University of Electronic Science and Technology of China, Chengdu 611731, P. R. China, and also with the Yangtze Delta Region Institute (Huzhou), University of Electronic Science and Technology of China, Huzhou 313001, P. R. China (e-mail:{liangyc@ieee.org}).}
\thanks{Y. Pei is with the Singapore Institute of Technology, 138683, Singapore (e-mail: yiyang.pei@singaporetech.edu.sg).}
\vspace{-0.3cm}
}
\maketitle

\IEEEpubid{
	\begin{minipage}{\textwidth}
		\centering
		\vspace{0.8in} 
		\footnotesize
		{© 2023 IEEE. Personal use of this material is permitted. Permission from IEEE must be obtained for all other uses, in any current or future media, including reprinting/republishing this material for advertising or promotional purposes, creating new collective works, for resale or redistribution to servers or lists, or reuse of any copyrighted component of this work in other works. 
		DOI: \href{http://doi.org/10.1109/TWC.2023.3286395}{10.1109/TWC.2023.3286395}}
	\end{minipage}
}

\begin{abstract}
Symbiotic radio (SR) is a promising solution to achieve high spectrum- and energy-efficiency due to its spectrum sharing and low-power consumption properties, in which the secondary system achieves data transmissions by backscattering the signal originating from the primary system. In this paper, we are interested in the pilot design and signal detection when the primary transmission adopts orthogonal frequency division multiplexing (OFDM). In particular, to preserve the channel orthogonality among the OFDM sub-carriers, each secondary symbol is designed to span an entire OFDM symbol. The comb-type pilot structure is employed by the primary transmission, while the preamble pilot structure is used by the secondary transmission. With the designed pilot structures, the primary signal can be detected via the conventional methods by treating the secondary signal as a part of the composite channel, i.e., the effective channel of the primary transmission. Furthermore, the secondary signal can be extracted from the estimated composite channel with the help of the detected primary signal. The bit error rate (BER) performance with both perfect and estimated CSI, the diversity orders of the primary and secondary transmissions, and the sensitivity to symbol synchronization error are analyzed. Simulation results show that the performance of the primary transmission is enhanced thanks to the backscatter link established by the secondary transmission. More importantly, even without the direct link, the primary and secondary transmissions can be supported via only the backscatter link.
\end{abstract}


\section{Introduction}

Trillions of Internet-of-Things (IoT) devices will be connected to the future sixth-generation (6G) network in order to support the immersive, intelligent, and ubiquitous services \cite{6GSCIS}. Such massive connections call for a large amount of spectrum and energy resources. Symbiotic radio (SR) is a promising technology to support this demand due to its spectrum sharing and passive transmission nature \cite{SRWC, SRTCCN, SRLong, PIEEE}. Specifically, there coexist two types of data transmissions in SR: the primary transmission and the secondary transmission \cite{SRTCCN, SRLong}. In the primary transmission, the primary transmitter (PTx) transmits messages with an active radio frequency (RF) chain by using the allocated dedicated spectrum. Sharing the spectrum of the primary transmission, the secondary transmitter (STx) modulates its information over the incident primary signal by changing its antenna load impedance periodically, and backscatters it to the cooperative receiver (CRx). Both the primary and secondary signals are detected at CRx. During this process, the secondary transmission can be achieved without dedicated spectrum and high power consumption for communications. Meanwhile, it provides beneficial multipath instead of harmful interference to the primary transmission, thus in return enhancing the performance of the primary transmission. In \cite{Mutualism}, the authors study the fundamental mutualistic mechanism between the primary and secondary transmissions. The conditions on the symbol period ratio between the primary and secondary signals are obtained to enable symbiosis in SR. In \cite{CapacitySR}, the capacity of multiple-input multiple-output (MIMO) SR is investigated, where the reflection behaviors of STx and the transmission of PTx are jointly optimized. 
Meanwhile, SR has been widely investigated in different scenarios to enable passive data transmission and enhancement to existing communication systems \cite{smallcell, UAVSR, fullduplex, NomaSR}. In \cite{smallcell}, STx assists the data transmission of the macro cell and simultaneously serves the micro users as a small cell base station (BS). In \cite{UAVSR}, an unmanned aerial vehicle (UAV) is leveraged to empower STx to transmit its collected environmental information to BS, and meanwhile attains the assistance from the secondary transmission.

Coherent receivers are typically used at CRx for jointly detecting the primary and secondary signals in SR. In \cite{CABC}, the optimal maximum-likelihood (ML) detector and suboptimal detectors have been proposed. In \cite{NovalParadigm}, the maximal-ratio-combining (MRC) based and successive-interference-cancellation (SIC) based detections are investigated for various SR systems. The active beamforming at PTx and the passive beamforming at STx are jointly optimized to minimize the bit error rate (BER) of the secondary transmission. In \cite{SRSM}, the symbiotic spatial modulation is proposed to help the receiver to detect the primary and secondary signals, in which STx requires the index information of the receiving antennas. In \cite{MPReceiverforSR}, the authors propose a transmission protocol for the SR system with multi-STx. The message-passing based receiver is investigated to jointly estimate the channel response, the primary and secondary signals. All the above studies assume the availability of full channel state information (CSI) at CRx, or require comprehensive collaboration between PTx and STx, which introduces significant overhead and cost to SR and especially the cost-sensitive STx.

On the other hand, since the secondary transmission is achieved via backscattering the primary signal, the primary and secondary signals are mathematically multiplied in the backscatter link \cite{washington}. In the case without the direct link from PTx to CRx, it is difficult for CRx to directly decouple the primary and secondary signals for the detection purpose due to the signal multiplications in the backscatter link. However, most existing receiver designs do not consider this scenario, and may fail to work when the direct link is blocked.

Orthogonal frequency division multiplexing (OFDM) has been widely applied in wireless communication systems, such as digital audio/video broadcasting (DAB/DVB) and Wi-Fi \cite{ofdmsurvey}. Hence, the extensively distributed OFDM systems are available as the primary component in SR, and can share their dedicated spectrum with the secondary transmission. The pilot design for OFDM includes the block-type structure and comb-type structure \cite{survey, channelestimationofdm}. Specifically, the block-type structure inserts pilot signals at all the sub-carriers periodically, while the comb-type structure inserts pilot signals at fixed sub-carriers in each OFDM symbol. When the cooperation is limited between the primary and secondary transmissions, the conventional receiver of SR over OFDM carriers extracts the secondary signal by measuring the changes of the effective channel with the help of the block-type pilot structure\cite{wifibackscatter}. Since it requires each secondary symbol to span a whole OFDM data frame, the block-type pilot design significantly limits the secondary data rate. To overcome this shortcoming, the comb-type pilot design can be helpful when the secondary data rate is much higher. Correspondingly, the pilot structure and receiver design for SR over OFDM carriers need to be further investigated.

To solve the above challenges, we are interested in the pilot design and signal detection for SR in this paper. In particular, we consider the case that the primary transmission adopts OFDM. The main contributions of this paper are summarized as follows:
\begin{itemize}
	\item We first establish the system model of SR over OFDM carriers, and then propose the pilot structure design for both primary and secondary transmissions. Specifically, each secondary symbol is designed to span an entire OFDM symbol to preserve the channel orthogonality. The comb-type pilot design is employed by the primary transmission, while the preamble pilot design is used by the secondary transmission.
	\item Based on the pilot structure design, we propose the signal detection methods of the primary and secondary signals utilizing estimated CSI. Specifically, by treating the secondary signal as a part of the composite channel (i.e., the effective channel of the primary transmission), the primary signal can be detected with conventional channel estimation and signal detection methods. The secondary signal can be extracted after estimating the direct and backscatter link CSI using the preamble pilots together with the detected primary signals. Notably, the secondary transmission in our proposed system is transparent to PTx, which means PTx does not need to change its transmitter or introduce any additional signaling overhead in order to accommodate the secondary transmission. Meanwhile, the detection of the primary and secondary signals is decoupled even in the case without any direct link.
	\item Then, we investigate the system performance with our proposed scheme, including the BER performance with both perfect and estimated CSI, the diversity orders of the primary and secondary transmissions, and the sensitivity to symbol synchronization error.
	\item Finally, simulation results show that a BER performance gain for the primary transmission can be achieved by exploiting the beneficial backscatter link. More importantly, the primary and secondary signals can still be recovered through the backscatter link even when the direct link is blocked. Additionally, our proposed SR system does not require the secondary signals to have perfect symbol synchronization with the primary ones, and it can tolerate some levels of symbol synchronization errors.
\end{itemize}

\emph{Organizations: }In Section \ref{system_model}, we establish the SR system model over OFDM carriers. In Section \ref{tranceiverdesign}, we propose the pilot structure design and the corresponding signal detection methods for the primary and secondary signals. In Section \ref{performanceanalysis}, the BER performance with both perfect and estimated CSI, the diversity orders of the primary and secondary transmissions, and the sensitivity to symbol synchronization error are analyzed. In Section \ref{simulation}, simulation results are presented to evaluate the performance of our proposed system. Finally, Section \ref{Conclusions} concludes this paper.

\emph{Notations:} The lowercase and boldface lowercase letter $a$ and $\ba$ denotes a scalar variable and vector, respectively. The boldface uppercase letter $\bA$ denotes a vector or a matrix. The uppercase calligraphic letter $\calA$ denotes a discrete and finite set. $|\cdot|$ means the operation of taking the absolute value if applied to a complex number, or the cardinal number if applied to a set. $\bbC^{a\times b}$ denotes the space of $a\times b$ complex-valued matrices. $\|\ba\|_2$ denotes the $l_2$-norm of vector $\ba$. $\bbE[a]$ denotes the statistical expectation of $a$. $\odot$ denotes the Hadamard product. $a ^ \dagger$ denotes the conjugate of the scalar $a$. $\bA^{-1}$, $\bA^\rmT$, and $\bA^\rmH$ denotes the inverse, transpose, and conjugate transpose of the matrix $\bA$, respectively. $\diag(\ba)$ returns a diagonal matrix whose diagonal elements are included in $\ba$. $\Re\{a\}$ denotes the real-part operation. $\bI_{a}$ denotes an identity matrix of size $a \times a$. $\calC\calN(\boldsymbol{\mu}, \boldsymbol{\Sigma})$ denotes the circularly symmetric complex Gaussian distribution with mean vector $\boldsymbol{\mu}$ and covariance matrix $\boldsymbol{\Sigma}$.

\section{System Model}\label{system_model}

\begin{figure}[t!]
	\centering
	\includegraphics[width=0.85\columnwidth]{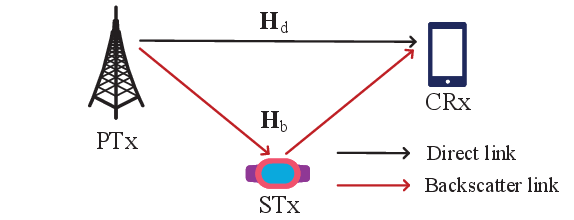}
	\caption{SR system model over OFDM carriers.}
	\label{systemmodel}
	\vspace{-0.5cm}
\end{figure}

\noindent As illustrated in Fig.~\ref{systemmodel}, we consider an SR system consisting of a single-antenna PTx, a single-antenna STx, and a single-antenna CRx\footnote{The results in this paper can be easily extended to the multi-antenna case.}. PTx, which uses an active RF chain, adopts OFDM to transmit the primary signal to CRx. STx periodically switches its reflection coefficient to modulate its secondary information over the incident primary signal and backscatters it to CRx. CRx needs to jointly detect the primary and secondary signals.

Denote $N$ as the number of OFDM sub-carriers.
Let $\bs(n) \triangleq [S_0(n), ..., S_k(n),..., S_{N - 1}(n)]^\rmT \in \bbC ^ {N \times 1}$ denote the block of the primary information signals transmitted during the $n$-th OFDM symbol, where the element $S_k(n)$ represents the signals over the $k$-th sub-carriers. It is assumed that $S_k(n)$, $k=0,..., N-1$ is modulated by $M_s$-ary quadrature amplitude modulation (QAM) with unit average power, i.e., $\bbE[|S_k(n)|^2] = 1$. Let $\calA_s$ be the modulation alphabet of the primary information signal, thus $S_k\left(n\right) \in \calA_s$. At PTx, $\bs(n)$ is first transformed to the time domain with the $N$-point inverse discrete Fourier transform (IDFT), which can be written as follows
\begin{align}
	\bx\left(n\right) = \frac{1}{\sqrt{N}}\bW^\rmH \bs(n),
\end{align}
where $\bW$ is the $N$-point DFT matrix whose $(p,q)$-th element is $e^{\frac{-j2\pi}{N}(p-1)(q-1)}$ for $p, q = 1,...,N$. Then, a cyclic prefix (CP) of length $N_{\rm cp}$ is appended to avoid inter-block interference (IBI). 

We assume that STx adopts $M_c$-ary phase-shift keying (PSK) to modulate its secondary information\footnote{It is demonstrated that high-order PSK modulation can be achieved by the backscatter phase modulator embedded in STx in \cite{backfi}.}. Denote $c(n) \in \calA_c$ the $n$-th transmitted secondary signal at STx, where $\calA_c$ represents its modulation alphabet. It is assumed that the symbol period of the secondary signal is the same as that of the primary OFDM symbol. This means each secondary signal spans an entire OFDM symbol. Furthermore, we assume that STx can achieve perfect synchronization with the OFDM symbol. In practice, this synchronization can be achieved and maintained with the help of a known synchronization sequence and the repeating CP structure of the primary signal\cite{modulationintheair, LTEbackscatter}. For the case with imperfect symbol synchronization, the tolerable range of symbol synchronization error will be analyzed in Section \ref{analysis_of_sync}.

Denote the channel impulse response (CIR) of the PTx-CRx link, the PTx-STx link, and the STx-CRx link by $\bh_\rmd$, $\bb$ and $\bg$, respectively. For convenience, the PTx-CRx link is referred to as the direct link, and the cascaded link from PTx to CRx via STx as the backscatter link. We assume that the PTx-CRx link, the PTx-STx link, and the STx-CRx link are all frequency-selective fading channels, with the corresponding number of equally spaced time domain taps denoted as $L_\rmd$, $L_1$, and $L_2$, respectively. Thus, the number of taps for the backscatter-link CIR is given as $L_\rmb = L_1 + L_2 -1$ due to the convolution of $\bb$ and $\bg$. We further denote $\bh(n)$ as the CIR of the composite link which combines the direct link and the backscatter link multiplied by $c(n)$. The number of taps for this composite link is given by $L = \max\{L_\rmd, L_\rmb + d_\rmb\}$, where $d_\rmb$ is the discrete propagation delay of the backscatter signal\footnote{Since STx is generally deployed near PTx or CRx to avoid the severe backscatter path loss, the delay of the backscatter signal compared with the direct signal can be negligible in most scenarios. To take a more global approach to our proposed system, the delay is thus considered in this paper.}. We can write $\bh(n)$ as $\bh(n)= [h_0(n), ..., h_l(n), ..., h_{L - 1}(n)]^\rmT$, where $h_l(n)$ denotes the $l$-th tap of the composite-link CIR. Additionally, a block fading channel model is utilized, assuming that the channels of the direct- and backscatter-links  remain constant within each secondary data frame. These secondary data frames consist of a maximum of $N_{\max}$ secondary signals.

Denote the direct-link channel frequency response (CFR), the backscatter-link CFR, and the composite-link CFR by $\bH_\rmd \triangleq [H_{\rmd, 0},...,H_{\rmd, k},...,H_{\rmd, N-1}]^\rmT\in \bbC^{N\times1}$, $\bH_\rmb \triangleq [H_{\rmb, 0}, ..., H_{\rmb, k},..., H_{\rmb, N - 1}]^\rmT \in \bbC^{N\times1}$, and $\bH(n) \triangleq [H_0(n), ...,H_k(n),..., H_{N - 1}(n)]^\rmT \in \bbC^{N\times1}$,
where $H_{\rmd, k}$, $H_{\rmb, k}$, and $H_k(n)$ denote the corresponding CFR at the $k$-th sub-carrier, respectively. By defining the matrix consisting of the first $L$ columns of $\bW$ as $\bF_L\in\bbC^{N\times L}$, the composite-link CFR can be derived from the composite-link CIR as
\begin{align}
	\bH\left(n\right) = \bF_L \bh\left(n\right).
\end{align}
Similarly, the direct- and backscatter-link CFR can be derived from the corresponding CIR by $\bH_\rmd = \bF_L\bar{\bh}_\rmd$ and $\bH_\rmb = \left(\bF_L\bar{\bb} \right)\odot \left(\bF_L\bar{\bg}\right)$, respectively, where $\bar{\bh}_\rmd$, $\bar{\bb}$, and $\bar{\bg}$ denote the zero-padded PTx-CRx link CIR of $\bh_\rmd$, the zero-padded PTx-STx link CIR of $\bb$, and the zero-padded STx-CRx link CIR of $\bg$ with the zero padding length of $L - L_\rmd$, $L - L_1$, and $L - L_2$, respectively.

We assume that the CP length $N_{\rm cp}$ is larger than the number of taps for the composite link $L$. Upon receiving the signal, CRx will first remove the CP and perform the $N$-point DFT. Then, the received signal of CRx in the frequency domain $\bY(n) \triangleq [Y_0(n), .., Y_k(n), ..., Y_{N - 1}(n)]^\rmT\in\bbC^{N\times1}$ can be written as
\begin{align}\label{receivedsignal}
	\bY\left(n\right) &= \sqrt{P_\rmT}\bS\left(n\right)\left(\bH_\rmd + c\left(n\right)\bH_\rmb\right) + \bU\left(n\right) \nonumber \\
	&= \sqrt{P_\rmT}\bS\left(n\right)\bH\left(n\right)+\bU\left(n\right),
\end{align}
where $P_\rmT$ is the transmission power, $\bS(n) = \diag(\bs(n))$, $\bH\left(n\right) = \bH_\rmd + c\left(n\right)\bH_\rmb$ is the composite-link CFR, and $\bU\left(n\right) \!\triangleq\! [U_0\left(n\right),...,U_k(n), ..., U_{N-1}\left(n\right)]^\rmT \sim\calC\calN(\mathbf{0}, \sigma^2\bI_N)$ is the frequency domain additive white Gaussian noise (AWGN). Here, $Y_k(n)$ and $U_k(n)$ denote the received signal and AWGN at the $k$-th sub-carrier. 

We have four observations from (\ref{receivedsignal}).
First, the primary signal experiences single-tap channels in the frequency domain thanks to the channel orthogonality of OFDM. Second, different from the conventional OFDM channels where the CFR remains constant over several OFDM symbols, the composite-link CFR in our system may change from one OFDM symbol to another due to the embedding of the secondary signal. Third, even when the direct link is blocked, the composite link still exists to carry the primary signal. Hence, the primary signal can still be recovered in such a case. Last, $c(n)$ is transmitted over all sub-carriers. Thus, there exists a spreading gain as can be seen in (\ref{perfectBERc}), (\ref{snrc1}), and (\ref{snrc2}) in Section \ref{performanceanalysis}.
\section{Pilot Structure and Receiver Design}\label{tranceiverdesign}
\noindent In this section, we will first present our proposed pilot structure, which is used for channel estimation. Then, based on the proposed pilot structure, we design the receiver signal detection methods to first detect the primary signal followed by the secondary signal.

\subsection{Pilot Structure}

\begin{figure}[t!]
	\centering
	\subfigure[Comb-type pilot design for the primary transmission.]{
		\begin{minipage}{0.9\columnwidth}
			\label{fig:Fig2a}	\includegraphics[width=1\columnwidth]{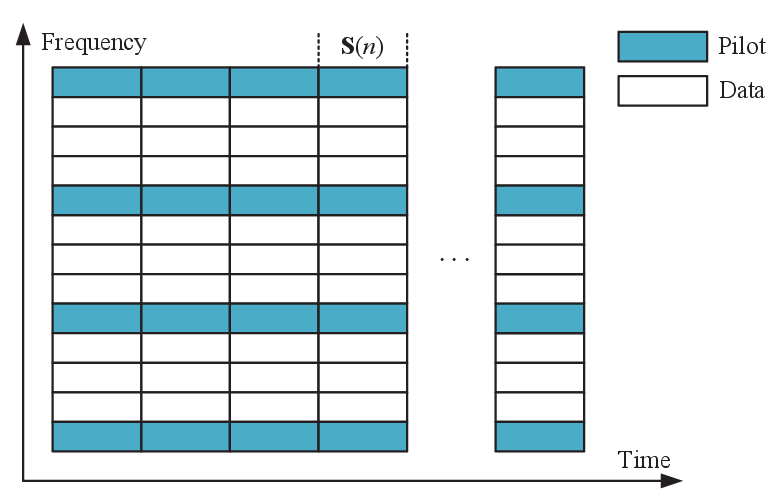}
		\end{minipage}
		}
	\subfigure[Preamble pilot design for the secondary transmission.]{
	\begin{minipage}{0.9\columnwidth}
		\label{fig:Fig2b}
		\includegraphics[width=1\columnwidth]{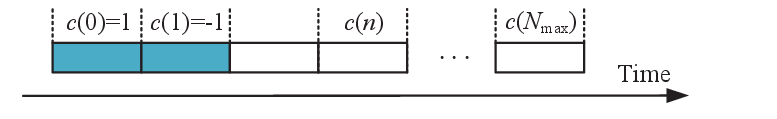}
	\end{minipage}
	}
	\vspace{-0.2cm}
	\caption{Pilot structure of SR system over OFDM carriers.}
	\vspace{-0.5cm}
	\label{fig:Fig2}
\end{figure}

\noindent Since each secondary signal spans an entire OFDM symbol, the channel orthogonality among the OFDM sub-carriers is preserved. Since the composite-link CFR $\bH\left(n\right)$ varies in each OFDM symbol due to the secondary signal $c\left(n\right)$, the block-type pilot design, where pilot signals are inserted periodically in all sub-carriers, does not work. In this paper, we adopt the comb-type pilot design for the primary transmission, which transmits pilot signals at fixed sub-carriers in each OFDM symbol. In this way, we can track the variation of the composite-link channel in each OFDM symbol.

As illustrated in Fig.~\ref{fig:Fig2a}, we consider that there are $N_\rmp$ sub-carriers carrying pilot signals in each OFDM symbol while the other $(N - N_\rmp)$ sub-carriers carrying primary data signals. By treating the secondary signal as a part of the composite channel, $\bH\left(n\right)$ can be estimated with the help of the comb-type pilots in each OFDM symbol. 

On the other hand, for the purpose of the secondary signal detection, CRx needs to acquire the CSI for both the direct and the backscatter links. To do so, we employ fixed preamble pilots in each frame of secondary signals. Specifically, two pilot symbols, which are respectively $c(0) = 1$ and $c(1)= -1$, are assigned at the beginning of each secondary data frame to separately estimate the direct-link CFR $\bH_\rmd$ and the backscatter-link CFR $\bH_\rmb$. As a result, the composite-link CFRs corresponding to the two pilot symbols can be written as $\bH(0) = \bH_\rmd + \bH_\rmb$ and $\bH(1) = \bH_\rmd - \bH_\rmb$. Thus, $\bH_\rmd$ and $\bH_\rmb$ can be estimated accordingly. 

It is noted that although we use the pilot symbols of $1$ and $-1$ as an example, they are not limited to these values. For the preamble pilot design of the secondary transmission, we have the following theorem.
\begin{mythe}\label{theorem1} 
	Leveraging $T$ $(T\ge2)$ pilot symbols arranged for the secondary transmission, the minimum variance of the estimated direct- and backscatter-link CFRs can be achieved when the pilot symbols satisfy $\sum_{n = 0}^{T - 1}c(n) = 0$ and $\left|c(n)\right|^2 = 1, n = 0, ..., T - 1$. With the estimated composite-link CFR $\hatbH(n)$, the estimation of the direct- and backscatter-link CFRs can be respectively derived as 
	\begin{align}\label{estimationwithT}
		\hatbH_\rmd = \frac{1}{T}\sum_{n = 0}^{T - 1} \tilbH(n), \quad \hatbH_\rmb = \frac{1}{T}\sum_{n = 0}^{T - 1} c^\dagger(n)\tilbH(n).
	\end{align}
	
\end{mythe}

\begin{IEEEproof}
	Please see Appendix \ref{theorem1proof}.
\end{IEEEproof}

\subsection{Primary Signal Detection}\label{primarytranceiver}

\noindent From (\ref{receivedsignal}), to detect the primary signal, we need to estimate the composite-link CFR $\bH\left(n\right)$.
We assume that $N_\rmp$ is equal to or larger than the number of taps for the composite link $L$. Utilizing the pilot-based channel estimation method, we first estimate the composite-link CIR $\bh\left(n\right)$. Then we can derive the composite-link CFR as $\bH\left(n\right) = \bF_L \bh\left(n\right)$. Finally, we can detect the primary signals by using the single-tap equalizer. 

Specifically, denote the $k$-th row of $\bF_L$ by $\boldf_k^\rmH$. By defining the matrix $\bF_\rmp \triangleq [\boldf_{k_1}, .., \boldf_{k_\rmp}]^\rmH\in\bbC^{N_\rmp \times L}$ where $k_1,...,k_\rmp$ are the pilot sub-carriers indices, the composite-link CFR at the pilot sub-carriers can be written as $\bH_\rmp\left(n\right) = \bF_\rmp\bh\left(n\right)\in\bbC^{N_\rmp\times 1}$. With the known pilot signals inserted in each OFDM symbol, the received signal at the pilot sub-carriers is expressed as
\begin{equation}
	\begin{split}\label{receivedsignalatpilot}
		\bY_\rmp\left(n\right) & = \sqrt{P_\rmT}\bS_\rmp \bH_\rmp\left(n\right) + \bU_\rmp\left(n\right) \\
		&= \sqrt{P_\rmT}\bS_\rmp\bF_\rmp\bh\left(n\right) + \bU_\rmp\left(n\right),
	\end{split}
\end{equation}
where $\bS_\rmp \in \bbC^{N_\rmp\times N_\rmp}$ is a diagonal matrix, the diagonal elements of which are the pilot signals, and $\bU_\rmp\left(n\right)\in\bbC^{N_\rmp\times 1}$ is the AWGN at the pilot sub-carriers. We assume the pilot signals are selected with the unit modulus, i.e., $\bS_\rmp^\rmH\bS_\rmp = \bI_{N_\rmp}$. Thus, $\bh\left(n\right)$ can be estimated as 
\begin{align}\label{LSEestimation}
	\tilbh\left(n\right) = \bG_0 \bY_\rmp \left(n\right),
\end{align}
where $\bG_0 = \left(P_\rmT\bF_\rmp^\rmH\bF_\rmp\right)^{-1}\sqrt{P_\rmT}\bF_\rmp^\rmH\bS_\rmp^\rmH$ is the least-square-error (LSE) estimator \cite{LSEestimateCIR}. After that, the composite-link CFR can be derived from the estimated CIR $\tilbh\left(n\right)$ as
\begin{align}\label{estimation0}
	\tilbH\left(n\right) = \bF_L\tilbh\left(n\right),
\end{align}
where $\tilbH(n) \triangleq [\tilH_0(n), ..., \tilH_k(n), ..., \tilH_{N - 1}(0)]^\rmT$, and $\tilH_k(n)$ represents the estimation of the composite-link CFR at the $k$-th sub-carrier.

Since the primary signal experiences the single-tap channel in the frequency domain as presented in (\ref{receivedsignal}), the primary signal at the $k$-th sub-carrier can be easily detected with the help of $\tilH_k(n)$ as follows
\begin{align}\label{decodes}
	\hatS_k(n) = \arg \min_{S\in\calA_s}\left|\frac{\tilH_k^\dagger(n)}{|\tilH_k(n)|^2}Y_k(n) - S\right|.
\end{align}
Note that the primary signal is detected without requiring the knowledge of the secondary signal, which means the secondary transmission is transparent to the primary transmission.
\subsection{Secondary Signal Detection} \label{2c}
\noindent Since the secondary transmission is achieved by passive backscattering at STx, the secondary signal $c\left(n\right)$ becomes a part of the composite-link CFR $\bH\left(n\right)$ as follows
\begin{align}
	\bH\left(n\right) = \bH_\rmb c\left(n\right) + \bH_\rmd.
\end{align}
Thus, $c\left(n\right)$ can be estimated as follows
\begin{align}\label{cdecoding}
	\hatc\left(n\right) = \arg\min_{c\in \calA_c}\left|\frac{\hatbH_\rmb^\rmH}{\|\hatbH_\rmb\|_2^2}\left(\hatbH\left(n\right) - \hatbH_\rmd\right) - c \right|,
\end{align}
where $\hatbH(n)$, $\hatbH_\rmd$, and $\hatbH_\rmb$ are the estimations of the composite-, direct-, and backscatter-link CFRs, respectively. From (\ref{cdecoding}), we notice that the composite-, direct-, and the backscatter-link CFRs need to be estimated first to detect the secondary signal. Thus, precise channel estimation is crucial for the detection of $c\left(n\right)$. 

Even though the estimation of the composite-link CFR $\tilbH(n)$ has been obtained before detecting $\hatbS(n)$, such estimation is achieved with only the pilot signals at the $N_\rmp$ sub-carriers. To improve the accuracy of the channel estimation, the detected primary signals $\hatbS\left(n\right)$ at all the $N$ sub-carriers can be reused for the channel estimation. Here, we provide two methods to estimate $\bH\left(n\right)$. One is directly performed in the frequency domain, and the other is operated in the time domain, utilizing the correlations among the sub-carriers.

\subsubsection{Method 1}
With the detected primary signals $\hatbS\left(n\right)$, the channel estimation can be performed directly in the frequency domain with the received signal $\bY\left(n\right)$. Utilizing the LSE estimator, the estimated composite-link CFR is obtained as
\begin{align}\label{estimation1}
	\hatbH\left(n\right) = \bG_1\bY\left(n\right),
\end{align}
where $\bG_1 = (\hatbS^\rmH(n) \hatbS(n))^{-1}\hatbS^\rmH\left(n\right) / \sqrt{P_\rmT}$ is the LSE estimator as a diagonal matrix.

\subsubsection{Method 2}
Similar to the channel estimation in Section \ref{primarytranceiver}, the composite-link channel can be firstly estimated in the time domain and then transformed to the frequency domain\cite{LSEestimateCIR}. Utilizing the LSE estimator, the estimated composite-link CFR can be obtained as
\begin{align}\label{estimation2}
	\hatbH\left(n\right) = \bF_L\bG_2\bY\left(n\right),
\end{align}
where $\bG_2 = (\bF_L^\rmH\hatbS^\rmH(n) \hatbS(n)\bF_L)^{-1}\bF_L^\rmH\hatbS^\rmH\!\left(n\right) / \sqrt{P_\rmT}$. For the case that the primary signals have the unit modulus, i.e., $\hatbS^\rmH(n)\hatbS(n) = \bI_N$, the estimator is simplified as $\bG_2 = \bF_L^\rmH\hatbS^\rmH\!\left(n\right) \!/ (N\sqrt{P_\rmT})$.

\begin{remark} \label{remark1}
	Since the correlation of the elements in $\bH\left(n\right)$ is considered in \emph{Method 2}, the performance of \emph{Method 2} is expected to be better than that of \emph{Method 1}, which will result in a performance gain for the detection of $c\left(n\right)$. On the other hand, the computation complexity of \emph{Method 1} is denoted by $\calO(N)$ since the diagonal matrix inversion can be directly derived. For the general case, the computational complexity of \emph{Method 2} mainly depends on the matrix inversion, and thus is denoted by $\calO(L^3 + L^2 N)$, which is higher than \emph{Method 1}. Thus, \emph{Method 1} can be adopted when the computation capacity of the secondary transmission is limited. Otherwise, \emph{Method 2} can be used for better performance of channel estimation.
\end{remark}

With the knowledge of $\hatbH\left(0\right)$ and $\hatbH\left(1\right)$, $\bH_\rmd$ and $\bH_\rmb$ can be estimated accordingly. Recall the two pilot symbols allocated at the beginning of the secondary data frame, which are $c\left(0\right) = 1$ and $c\left(1\right) = -1$. The composite-link CFR of the pilot symbols are $\bH\left(0\right) = \bH_\rmd + \bH_\rmb$ and $\bH\left(1\right) = \bH_\rmd - \bH_\rmb$, respectively. Thus, we can estimate $\bH_\rmd$ and $\bH_\rmb$ as follows
\begin{subequations}\label{estimationdb}
\begin{align}
	\hatbH_\rmd &= \frac{1}{2}\left(\hatbH\left(0\right) + \hatbH\left(1\right)\right), \\
	\hatbH_\rmb &= \frac{1}{2}\left(\hatbH\left(0\right) - \hatbH\left(1\right)\right).
\end{align}
\end{subequations}
Once $\hatbH_\rmd$ and $\hatbH_\rmb$ are obtained, we can extract $c\left(n\right)$ from $\hatbH\left(n\right)$ based on (\ref{cdecoding}).

\subsection{Overall Detection}

\noindent Based on the description above, we summarize the joint signal detection as follows. With the received frequency-domain samples $\bY(n)$, we first estimate the composite-link CFRs $\tilbH(n)$ and then use it to detect the primary signals $\hatbS(n)$. Next, based on the detected primary signals, the composite-link CFRs are estimated again via \emph{Method 1} or \emph{Method 2}. After that, the direct- and backscatter-link CFRs $\hatbH_\rmd$ and $\hatbH_\rmb$ are estimated. Finally, the secondary signals $\hatc(n)$ are detected.

\begin{algorithm}[htbp]
	\caption{Joint primary and secondary signal detection.}
	\label{Algorithm1}
	\begin{algorithmic}[1]
		\REQUIRE The received frequency-domain samples $\bY(n)$, $\forall n \in \{0, 1, ...,N_{\max} - 1\}$.\\
		\ENSURE The primary signals $\hatbS(n)$ and the secondary signals $\hatc(n)$, $\forall n \in \{0, 1, ...,N_{\max} - 1\}$.\\
		\FOR{$n = 0$ to $N_{\max} - 1$}
		\STATE Estimate $\tilbH(n)$ by (\ref{LSEestimation}) and (\ref{estimation0}).
		\STATE Detect $\hatbS(n)$ by (\ref{decodes}).
		\STATE Estimate $\hatbH(n)$ by (\ref{estimation1}) or (\ref{estimation2}).
		\ENDFOR
		\STATE Estimate $\hatbH_\rmd$ and $\hatbH_\rmb$ by (\ref{estimationdb}).
		\FOR{$n = 2$ to $N_{\max} - 1$}
		\STATE Detect $\hatc(n)$ by (\ref{cdecoding}).
		\ENDFOR
		\RETURN $\hatbS(n)$ and $\hatbc(n)$, $\forall n$.
	\end{algorithmic}
\end{algorithm}

\begin{remark}
	The block of primary signals in one OFDM symbol and the secondary signal are jointly detected by the optimal ML detector\cite{CABC}. The computation complexity of the two-step ML detector can be derived as $\calO(N M_c M_s)$. For the case with high-order modulation of the secondary transmission or a large number of OFDM sub-carriers, the arithmetic cost can be very high for CRx. As for our proposed method, the primary signals of one OFDM symbol can be detected first, followed by the detection of the secondary signals. Thus, the computation complexity of our proposed detection method can be derived as $\calO(M_c + N M_s)$. Together with the computation complexity of channel estimation in \emph{Remark} \ref{remark1}, the arithmetic cost of our proposed receiver is lower than the ML detector.
\end{remark}

\begin{figure}[t!]
	\centering
	\includegraphics[width=0.99\columnwidth]{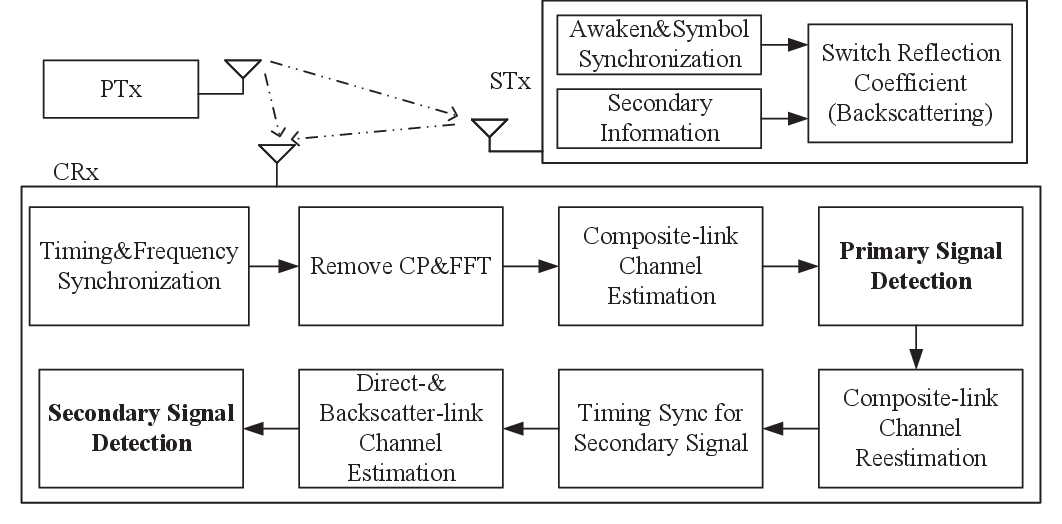}
	\caption{Brief block diagram of SR system over OFDM carriers.}
	\label{blockdiagram}
	\vspace{-0.5cm}
\end{figure}

The block diagram of our proposed system is shown in Fig.~\ref{blockdiagram}.
To provide a more detailed explanation of the system procedure, we consider each primary data frame to consist of two parts. The first one is the frame head, which includes a known sequence used for accurate timing synchronization to find the start of one data frame. The second one is the data component, which holds the primary signals for transmission.
In the previous sections, we consider the case that the data frames of the primary and secondary transmissions are aligned with each other, and timing synchronization is performed before signal detection. Nevertheless, the received secondary data frame can be delayed compared with the received primary data frame since STx is first awakened by the primary signal, and then decides to transmit its secondary data frame immediately or later. For the delayed case, the timing synchronization for the primary and secondary data frames can be performed separately as presented in Fig.~\ref{blockdiagram}. Specifically, after the signal reception at CRx, the timing synchronization for the primary data frame can be performed at the beginning with the conventional method, such as using the cross-correlation between the known sequence in the frame head and the received time-domain signals \cite{synchronization}. For the secondary data frame, the timing synchronization can be performed by the cross-correlation between the preamble pilots and the estimated CFRs $\bH(n)$, $\forall n$, after the composite-link channel re-estimation. Notice that the preamble pilot number of the secondary data frame needs to be increased to a number of more than two for better synchronization performance.

\section{Performance Analysis} \label{performanceanalysis}
\noindent In this section, we characterize the performance of the proposed system by analyzing the BER performance with both perfect and estimated CSI, the diversity orders of both the primary and secondary transmissions, and the sensitivity to symbol synchronization error. Without loss of generality, we consider that PTx and STx adopt $M_s$-ary QAM and $M_c$-ary PSK modulation schemes, respectively. Note that the analysis can be easily generalized to other modulation schemes. 

\subsection{BER Performance with Perfect CSI}
\noindent Recall that the primary signal detection relies on the CSI for the composite link, while the secondary signal detection requires the CSI for both the direct and backscatter links. Hence, we assume the CSI for the composite link is perfectly known for the primary signal detection, and the CSI for both the direct and backscatter links is perfectly known for the secondary signal detection\footnote{We note that this assumption of perfect CSI has no effect on the detection process of our proposed receiver. Even though this assumption implies perfect knowledge of secondary signals for the primary signal detection, detection errors may still occur for the secondary signal detection due to the noise term in the receiver signal. Moreover, it should be noted that the BER performance with perfect CSI indicates the best performance that our proposed receiver can achieve, regardless of the channel estimation methods adopted.}. Based on this assumption, the BER performance can be accordingly derived.
\subsubsection{Primary transmission}
Since each secondary symbol spans an entire OFDM symbol, the backscatter link can be regarded as a beneficial multipath component to the detection of $\bS\left(n\right)$. 
Given $\bH_\rmd$ and $\bH_\rmb$, the BER of the primary transmission can be derived as
\begin{multline}
	P_{e, s} = \frac{1}{|\calD|M_c}\sum_{k \in \calD}\sum_{c\in\calA_c}\left(1- \left(1 - 2 \left(1 - \frac{1}{\sqrt{M_s}}\right)\right.\right. \\ 
	\left.\left.\calQ\left(\sqrt{\frac{3P_\rmT\left|H_{\rmd, k} +  cH_{\rmb, k}\right|^2}{\left(M_s - 1\right)\sigma^2}}\right)\right) ^ 2\right),
\end{multline}
where the $\calQ$-function is $\calQ(z) = (1 / \sqrt{2\pi})\int_{z}^{\infty}e^{-u^2/2} du$, and $\calD$ is the set of the indices for the data sub-carriers. We can see that $P_{e, s}$ is dependent on the composite-link channel, which is relevant to the secondary signal.
\subsubsection{Secondary transmission}
In SR, the BER of the secondary transmission is theoretically coupled with the BER of the primary transmission. The exact BER expression of the secondary transmission involves nonlinear multivariable equations, which need to be solved by iterative-based method \cite{Mutualism}. For simplicity, we consider the case that the primary signal is perfectly detected, in which the result can be regarded as the lower bound of the BER of the secondary transmission. With a Gray code used in the $M_c$-PSK mapping, we can obtain the approximate BER \cite{DigitalComm} of the secondary transmission in the high SNR regime as
\begin{align}\label{perfectBERc}
	P_{e, c} \approx \frac{2}{\log_2 M_c}\calQ \left(\sqrt{2\sin^2\left(\frac{\pi}{M_c}\right)\frac{P_\rmT\left\|\bH_\rmb\right\|_2^2}{\Gamma_1\sigma^2}}\right),
\end{align}
where $\Gamma_1 = \bbE[|1 / S_k(n)|^2]$.
We notice that $P_{e,c}$ is dependent on the backscatter-link channel and that a stronger backscatter link will result in a lower BER for the secondary signal detection. Moreover, a spreading gain can be achieved by the combination operation over all the sub-carriers.

\subsection{BER Performance with Estimated CSI} \label{estimationwitherror}
\noindent Firstly, we can rewrite the estimation of the composite-link CFR $\bH_{\rm est}(n)$ by 
\begin{align}
    \bH_{\rm est}(n) = \bH(n) + \bepsilon(n),
\end{align}
where $\bH_{\rm est}(n)$ can be replaced by $\tilbH(n)$ or $\hatbH(n)$, and $\bepsilon(n)$ is the channel estimation error. To further analyze the effect of the channel estimation on the BER performance, we first derive the sufficient statistics for the detection of the primary and secondary transmissions, and then calculate the SNR with the channel estimation error, in order to obtain the BER performance with estimated CSI. 
\subsubsection{Primary transmission}
\noindent Denote the channel estimation error from (\ref{LSEestimation}) and (\ref{estimation0}) by $\bepsilon_0\left(n\right)$, which can be expressed as $\bepsilon_0\left(n\right) = \bF_L\bG_0\bU_\rmp\left(n\right)$. The $k$-th element of $\bepsilon_0\left(n\right)$ is distributed as $\epsilon_{0, k}(n)\sim \calC\calN(0, \sigma^2 \boldf_k^\rmH (\bF_\rmp^\rmH \bF_\rmp)^{-1}\boldf_k / P_\rmT)$.

The $k$-th element of $\tilbH(n)$ can be rewriten as $\tilH_k(n) = H_k(n) + \epsilon_{0, k}(n)$. From (\ref{decodes}), the primary signal at the $k$-th sub-carrier $S_k\left(n\right)$ is detected from the following sufficient statistic
\begin{multline}\label{dss1}
	\tilH_k^\dagger\left(n\right)Y_k\left(n\right) = \sqrt{P_\rmT} \left|H_k\left(n\right)\right|^2 S_k\left(n\right) + \\
	\sqrt{P_\rmT} \epsilon_{0, k}^\dagger\!\left(n\right) \! H_k\!\left(n\right) \! S_k\!\left(n\right) \!+\! \left(H_k\!\left(n\right) \!+\! \epsilon_{0, k}\!\left(n\right)\right)^\dagger U_k\!\left(n\right).
\end{multline}
Since the channel estimation is performed in each OFDM symbol, $\epsilon_{0, k}\left(n\right)$ varies with $n$. We regard the first term in the right-hand side of (\ref{dss1}) as the desired signal and the other terms as the effective noise. Given the secondary signal $c$, the SNR of $S_k\left(n\right)$ is given by
\begin{align} \label{snrswithestimatedcsi}
\gamma_{s, k}\!(c) \!=\! \frac{P_\rmT\left|H_{\rmd, k} + cH_{\rmb, k}\right|^2}{\sigma^2\!\left(\boldf_k^\rmH \left(\bF_\rmp^\rmH \bF_\rmp\right)^{-1}\boldf_k \!+\! 1 \!+\! \frac{\sigma^2 \boldf_k^\rmH \left(\bF_\rmp^\rmH\bF_\rmp\right)^{-1}\boldf_k}{P_\rmT\left|H_{\rmd, k} + cH_{\rmb, k}\right|^2}\right)}.
\end{align}
When the pilot signals are allocated at equally spaced sub-carriers, we can obtain the fact that $\bF_\rmp^\rmH\bF_\rmp = N_\rmp \bI_{N_\rmp}$. And the SNR of $S_k(n)$ can be rewritten as 
\begin{align}
	\gamma_{s, k}(c) \!=\! \frac{P_\rmT\left|H_{\rmd, k} + cH_{\rmb, k}\right|^2}{\sigma^2\!\left(\frac{N_\rmp + L}{N_\rmp}\!+\! \frac{L \sigma^2}{N_\rmp P_\rmT\left|H_{\rmd, k} + cH_{\rmb, k}\right|^2}\right)}.
\end{align}
Comparing with the case of perfect CSI, we can notice that the SNR of $S_k(n)$ is degraded since the noise power is amplified by a positive number larger than one.
Meanwhile, the BER of the primary transmission with estimated CSI can be written as
\begin{multline}
	P_{e, s} = \frac{1}{|\calD|M_c}\sum_{k \in \calD} \sum_{c\in\calA_c}\left(1 - \left(1 - 2 \left(1 - \frac{1}{\sqrt{M_s}}\right)\right.\right. \\
	\left.\left.\calQ\left(\sqrt{\frac{3\gamma_{s, k}(c)}{M_s - 1}}\right)\right)^2\right).
\end{multline}
We notice that $P_{e,s}$ is dependent on the composite-link CFR, as well as the numbers of the pilot sub-carriers and taps for the composite link. A larger number of pilot sub-carriers $N_\rmp$ or a smaller number of taps for the composite link $L$ will lead to a lower BER.

\subsubsection{Secondary transmission}
\noindent We consider the BER of the secondary transmission with perfectly detected primary signal. 
We first analyze the BER of the secondary transmission via \emph{Method 1} as follows.
Denote the channel estimation error via \emph{Method 1} in (\ref{estimation1}) as $\bepsilon_1(n)$, which can be expressed as $\bepsilon_1(n) = \bG_1 \bU(n)$. The estimation error is distributed as $\bepsilon_1(n) \sim \calC\calN(\mathbf{0}, \sigma^2 (\bS^\rmH(n)\bS(n))^{-1} / P_\rmT)$.

From (\ref{estimationdb}), we can rewrite $\hatbH_\rmd$ and $\hatbH_\rmb$ as
\begin{subequations}\label{estimationerrordb}
\begin{align}
	\hatbH_\rmd & = \bH_\rmd + \frac{\bepsilon_1\left(0\right) + \bepsilon_1\left(1\right)}{2} = \bH_\rmd + \bepsilon_\rmd, \\
	\hatbH_\rmb & = \bH_\rmb + \frac{\bepsilon_1\left(0\right) - \bepsilon_1\left(1\right)}{2} = \bH_\rmb + \bepsilon_\rmb,
\end{align}
\end{subequations}
where $\bepsilon_\rmd \!=\! \left(\bepsilon_1\left(0\right) \!+\! \bepsilon_1\left(1\right)\right) / 2$ and $\bepsilon_\rmb \!=\! \left(\bepsilon_1\left(0\right) \!-\! \bepsilon_1\left(1\right)\right) / 2$. We observe that $\bepsilon_\rmd$ and $\bepsilon_\rmb$ are correlated with each other. Meanwhile, $\bepsilon_\rmd$ and $\bepsilon_\rmb$ are fixed for the detection of the subsequent $c(n)$ in each secondary data frame.

Recalling the detection of $c\left(n\right)$ described in (\ref{cdecoding}), $c\left(n\right)$ is detected from the following sufficient statistic
\begin{align}\label{dss2}
	&\hatbH_\rmb^\rmH\left(\hatbH\left(n\right) - \hatbH_\rmd\right) \nonumber\\ 
	= &\left(\bH_\rmb + \bepsilon_\rmb\right)^\rmH\left(\bH_\rmb c\left(n\right) + \bepsilon_1\left(n\right) - \bepsilon_\rmd\right)  \nonumber\\
	= &\left(\bH_\rmb + \bepsilon_\rmb\right)^\rmH\bH_\rmb c\left(n\right) + \left(\bH_\rmb + \bepsilon_\rmb\right)^\rmH\left(\bepsilon_1\left(n\right) - \bepsilon_\rmd\right).
\end{align}
Given $\bepsilon_\rmd$ and $\bepsilon_\rmb$, we can obtain the BER performance for one secondary data frame. By calculating the expectation of the BER concerning $\bepsilon_\rmd$ and $\bepsilon_\rmb$, the average BER with estimated CSI can be obtained as follows
\begin{align}\label{Expectation}
	P_{e, c}^{(1)} \!=\! \frac{1}{M_c}\bbE_{\bepsilon_\rmd, \bepsilon_\rmb}\!\!\!\left[\sum_{c\in\calA_c}\!\!\calQ\!\left(\frac{\left(\bH_\rmb \!+\! \bepsilon_\rmb\right)^\rmH\left(\bH_\rmb c \!-\! \bepsilon_\rmd\right)}{\sqrt{\frac{\sigma^2}{2P_\rmT}\!\!\left(\left\|\bH_\rmb\right\|_2^2 \!+\! \bepsilon_\rmb^\rmH\bepsilon_\rmb\right)}}\right)\right]\!\!.
\end{align}
However, it is challenging to derive the closed-form expression of (\ref{Expectation}).

We regard the first term in (\ref{dss2}) as the desired signal and the other terms as the effective noise. Unlike $\bepsilon_\rmd$ and $\bepsilon_1(n)$ which contribute to the effective noise, we notice that $\bepsilon_\rmb$ contributes to both the desired signal and the effective noise. In the high SNR regime, the contribution of $\bepsilon_\rmb$ to the desired signal can be negligible. Thus, we can obtain the following theorem. 
\begin{mythe}\label{theorem2}
Given $\bH_\rmb$, the SNR of $c\left(n\right)$ with estimated CSI via Method 1 can be expressed as
\begin{align}\label{snrc1}
	\gamma_c^{(1)}  = \frac{P_\rmT\left\|\bH_\rmb\right\|_2^2}{\sigma^2\left(2 \Gamma_1 + \frac{N(2 \Gamma_1 ^ 2 + \Gamma_2)\sigma^2}{4P_\rmT\left\|\bH_\rmb\right\|_2^2}\right)},
\end{align}
where $\Gamma_2 = \bbE[|1 / S_k(n)|^4]$. If the primary signals have a constant modulus, i.e., $|S_k(n)|^2 = 1$, the SNR can be simplified as
\begin{align}\label{snrc11}
	\gamma_c^{(1)} = \frac{P_\rmT\left\|\bH_\rmb\right\|_2^2}{\sigma^2\left(2 \Gamma_1 + \frac{3N\sigma^2}{4P_\rmT\left\|\bH_\rmb\right\|_2^2}\right)}.
\end{align}
\end{mythe}
\begin{IEEEproof}
Please see Appendix \ref{theorem23proof}.
\end{IEEEproof}
Comparing with the case of perfect CSI, we notice that the average power of the effective noise of (\ref{snrc1}) is enlarged by a factor of $\left(2 \Gamma_1 + \frac{N(2 \Gamma_1 ^ 2 + \Gamma_2)\sigma^2}{4P_\rmT\left\|\bH_\rmb\right\|_2^2}\right)$, which is a positive number larger than $2\Gamma_1\ge2$. 
This indicates that the SNR of $c(n)$ is degraded by the channel estimation error.
Furthermore, the BER of $c(n)$ with estimated CSI via \emph{Method 1} can be expressed as
\begin{align}
	P_{e, c}^{(1)} \approx \frac{2}{\log_2 M_c}\calQ\left(\sqrt{2\sin^2\left(\frac{\pi}{M_c}\right)\gamma_c^{(1)}}\right).
\end{align}

Next, we analyze the BER of the secondary transmission via \emph{Method 2} as follows.
Denote the channel estimation error via \emph{Method 2} in (\ref{estimation2}) as $\bepsilon_2\left(n\right)$, which can be expressed as $\bepsilon_2\left(n\right) = \bF_L\bG_2 \bU\left(n\right)$ and distributed as $\bepsilon_2\left(n\right)\sim\calC\calN\left(\mathbf{0}, \sigma^2 \bF_L(\bF_L^\rmH\bS^\rmH(n)\bS(n)\bF_L)^{-1}\bF_L^\rmH/NP_\rmT\right)$. From (\ref{estimationerrordb}), $\bepsilon_\rmd$ and $\bepsilon_\rmb$ can be derived by replacing $\bepsilon_1\left(n\right)$ with $\bepsilon_2\left(n\right)$, as well as the sufficient statistic in (\ref{dss2}). Since it is challenging to derive the closed-form of the matrix inversion of $(\bF_L^\rmH\bS^\rmH(n)\bS(n)\bF_L)^{-1}$, we assume that $\bS^\rmH(n)\bS(n) = \bI_N$ and derive the approximate SNR expression of $c(n)$. Similarly, we can obtain the following theorem.
\begin{mythe}\label{theorem3}
	Given $\bH_\rmb$, the SNR of $c\left(n\right)$ with estimated CSI via \emph{Method 2} can be expressed as
	\begin{align}\label{snrc2}
		\gamma_c^{(2)}  = \frac{P_\rmT\left\|\bH_\rmb\right\|_2^2}{\sigma^2\left(2 + \frac{3L\sigma^2}{4P_\rmT\left\|\bH_\rmb\right\|_2^2}\right)}.
	\end{align}
\end{mythe}
\begin{IEEEproof}
	Please see Appendix \ref{theorem23proof}.
\end{IEEEproof}
From (\ref{snrc11}) and (\ref{snrc2}), we notice that the detection performance of the secondary transmission is dependent on the backscatter-link CFR. Besides, the BER performance of \emph{Method 1} is also dependent on the number of OFDM sub-carriers $N$, and that of \emph{Method 2} is dependent on the number of taps for the composite link $L$. The BER performance of \emph{Method 2} is much better than that of \emph{Method 1} from the terms of the SNR due to the fact that $N \gg L$. Meanwhile, the BER of $c(n)$ with estimated CSI via \emph{Method 2} can be expressed as
\begin{align}\label{gamma12}
	P_{e, c}^{(2)} \approx \frac{2}{\log_2 M_c}\calQ\left(\sqrt{2\sin^2\left(\frac{\pi}{M_c}\right)\gamma_c^{(2)}}\right).
\end{align}
For the case of other preamble pilot arrangements which satisfy the conditions in Theorem \ref{theorem1}, $\bepsilon_\rmd$ and $\bepsilon_\rmb$ can be derived according to (\ref{estimationwithT}) and (\ref{estimationdb}). Furthermore, the results of the BER performance can be easily extended as in Appendix \ref{theorem23proof}.

\subsection{Diversity Order} \label{diversityorder}

\noindent To further analyze the diversity orders of the primary and secondary transmissions, we consider a stochastic channel description and analyze the average BER performance.

\subsubsection{Primary transmission}

Since the secondary transmission is transparent to the primary transmission and the channel coding is not considered in our proposed system, the primary transmission can be regarded as an uncoded OFDM transmission. The channel can have a null on or close to the OFDM sub-carrier. In other words, each sub-carrier channel may be in a deep fade, and the detection of the corresponding primary signal may fail \cite{tse}. Even though the probability of such failed detection may be relatively low for each sub-carrier, the loss of the diversity caused by such events can be significant when the channel exhibits frequency-selective fading. As a result, the uncoded OFDM has the diversity order of only one \cite{OFDM_single_carrier}. By adopting a precoder such as a linear precoder before the DFT operation at PTx, the diversity order of the primary transmission can be efficiently increased over Rayleigh fading channels with multiple taps \cite{LinearPrecoder}.

\subsubsection{Secondary transmission}
We consider the average BER of the secondary signal with perfect CSI and perfectly detected primary signal. Without loss of generality, we consider that $M_c = 2$ for simplicity. We assume that the channel taps of the backscatter link are equally spaced and Rayleigh fading with zero mean and equal power $\sigma_\rmb^2$. Denote the normalized backscatter-link CIR by $\bar{\bh}_\rmb = \bh_\rmb / \sigma_\rmb$. Hence, $\|\bar{\bh}_\rmb\|_2^2/\sigma_\rmb^2$ follows the Chi-square distribution with $2L_\rmb$ degrees of freedom, whose probability density function (PDF) is given by $\frac{1}{(L_\rmb - 1)!}x^{L_\rmb - 1}e^{-x}$ with $x\ge0$. Furthermore, the average BER of the secondary signal can be derived as
\begin{align}\label{averageBERc}
	\bar{P}_{e, c} \!&=\! \bbE_{\bH_\rmb}\!\!\left[\!\calQ\!\left(\!\sqrt{\frac{2P_\rmT\!\left\|\bH_\rmb\right\|_2^2}{\Gamma_1\sigma^2}}\right)\!\right]  \!\!\overset{(a)}{=}\! \bbE_{\bh_\rmb}\!\left[\!\calQ\!\left(\!\sqrt{\frac{2P_\rmT\!\left\|\bh_\rmb\right\|_2^2}{\Gamma_1\sigma^2}}\right)\!\right]\nonumber \\
	&= \bbE_{\bh_\rmb}\left[\calQ\left(\sqrt{\frac{2\gamma_\rmb\left\|\bar{\bh}_\rmb\right\|_2^2}{\sigma_\rmb^2}}\right)\right] \nonumber\\
	&= \int_{0}^{\infty}\calQ\left(\sqrt{2\gamma_\rmb x}\right) \frac{1}{\left(L_\rmb - 1\right)!}x^{L_\rmb - 1}e^{-x}dx \nonumber \\
	&\overset{(b)}{=} \left(\frac{1 - \mu}{2}\right)^{L_\rmb}\sum_{l = 0}^{L_\rmb - 1}{L_\rmb - 1 + l \choose l}\left(\frac{1 + \mu}{2}\right)^l \nonumber \\
	&\overset{(c)}{\approx} {2L_\rmb - 1 \choose L_\rmb	}\frac{1}{\left(4\gamma_\rmb\right)^{L_\rmb}},
\end{align}
where $\mu = \sqrt{\gamma_\rmb/(1 + \gamma_\rmb)}$, and $\gamma_\rmb = P_\rmT\sigma_\rmb^2 / (\Gamma_1\sigma^2)$ is the average backscatter link SNR. In (\ref{averageBERc}), $(a)$ holds from the fact that the channel gain is equal in both the time and frequency domains, and $(b)$ and $(c)$ hold based on \cite{DigitalComm}. 

From (\ref{averageBERc}), we can notice that $\bar{P}_{e, c}$ decreases significantly when the number of taps for the backscatter link $L_\rmb$ increases. With respect to $\gamma_\rmb$, we can derive the negative slope of the BER curve in the high SNR regime which is called diversity order as
\begin{align}
	D_c = -\lim_{\gamma_\rmb \rightarrow \infty} \frac{\log_{10}\bar{P}_{e, c}}{\log_{10}\gamma_\rmb} = L_\rmb.
\end{align}
Notice that the diversity order for the secondary transmission is $L_\rmb$, which means that the secondary transmission can obtain a frequency diversity gain from the backscatter link in our proposed SR system. This is because the secondary transmission can be regarded as a single-carrier transmission with repetition code, where each secondary signal $c(n)$ has $(N + N_{\rm cp})$ repetitions in the time domain. Since the backscatter-link channel has $L_\rmb$ taps, CRx can receive delayed replicas from $L_\rmb$ branches of diversity with respect to each repetition. Leveraging the structure of OFDM, the received repetitions with their replicas are transformed into the frequency domain for combination and detection. As a result, the repetition code brings an SNR gain due to the block-fading channel model. And the backscatter-link channel brings a frequency diversity gain of $L_\rmb$ for the secondary transmission.
For the case with estimated CSI, the approximate diversity order of $L_\rmb$ can be obtained for the secondary transmission, which can be validated in Section \ref{simulation}.

\subsection{Sensitivity to Symbol Synchronization Error} \label{analysis_of_sync}

\noindent Recall that perfect symbol synchronization between the OFDM symbol and the secondary signal is assumed at STx in Section \ref{system_model}. However, the distorted synchronization sequence at STx due to the potential deep fading of the PTx-STx link, and the hardware impairment may lead to symbol synchronization error. 
Such symbol synchronization errors can increase the number of taps for the composite link, and cause the composite-link channel to change within one OFDM symbol, which can affect the channel orthogonality among the OFDM sub-carriers. 
In addition, it can also lead to the failure of channel estimation for the composite-link channel, if the number of taps for the composite link exceeds that of the pilot sub-carriers. As a result, both primary and secondary signal detection can be affected. Even though symbol synchronization is crucial, our proposed system does not require absolute perfection of symbol synchronization between the primary and secondary signals. In the following, we will derive the tolerable range of symbol synchronization errors such that our proposed system can still work well. 

\begin{figure}[t!]
	\centering
	\includegraphics[width=0.85\columnwidth]{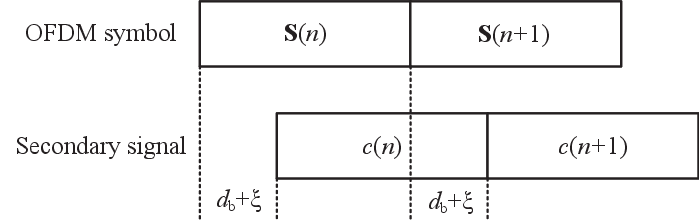}
	\caption{The received primary and secondary signals with the symbol synchronization error $\xi$.}
	\label{imperfect_sync}
	\vspace{-0.5cm}
\end{figure}

\begin{figure}[t!]
	\centering
	\includegraphics[width=0.5\columnwidth]{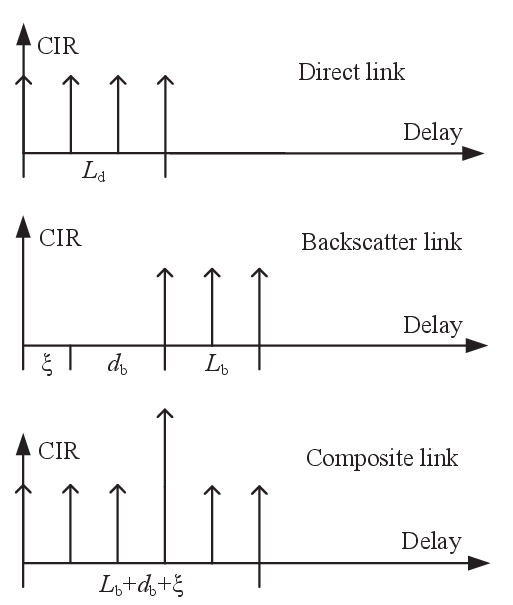}
	\caption{The direct-, backscatter-, and composite-link CIRs with symbol synchronization error $\xi$.}
	\label{sync_error}
	\vspace{-0.5cm}
\end{figure}

Firstly, we denote the discrete synchronization error between the OFDM symbol and the secondary signal by $\xi \in \left[0, N + N_{\rm cp} - 1\right)$ as illustrated in Fig.~\ref{imperfect_sync}, which means that each secondary symbol is delayed for $\xi$ discrete samples compared with the start of the corresponding OFDM symbol. As presented in Fig.~\ref{sync_error}, the number of taps for the composite link can be rewritten as $L = \max\left\{L_\rmd, L_\rmb + d_\rmb + \xi\right\}$. Recall that the CP length $N_{\rm cp}$ should be no less than the number of taps for the composite link to avoid the IBI, i.e., $N_{\rm cp} \ge L - 1$. Two cases are discussed for the value of $L_\rmd$. For the first case that $L_\rmd \ge L_\rmb + d_\rmb + \xi$, the synchronization error $\xi$ has no impact on the number of taps for the composite link $L$. As long as $N_{\rm cp} \ge L_\rmd - 1$, the channel orthogonality can remain. For the second case that $L_\rmd < L_\rmb + d_\rmb + \xi$, the channel orthogonality can be maintained when $N_{\rm cp} \ge L_\rmb + d_\rmb + \xi - 1$. As a result, the channel orthogonality among the OFDM sub-carriers can be preserved when $\xi \in [0, N_{\rm cp} - L_\rmb - d_\rmb + 1]$. Otherwise, the loss of the channel orthogonality can cause the IBI and inter-channel interference (ICI) \cite{ICIIBI}, and further affect the primary and secondary signal detection.

On the other hand, when $L_\rmd < L_\rmb + d_\rmb + \xi \le N_\rmp$, the increase of $\xi$ can enlarge the number of taps for the composite link $L$. Since the number of channel taps to be estimated increases, the performance of estimation in (\ref{LSEestimation}), (\ref{estimation1}), and (\ref{estimation2}) is correspondingly degraded. However, the composite-link CFR can still be estimated successfully. When $L_\rmb + d_\rmb + \xi > N_\rmp$, we notice that the matrix $\sqrt{P_\rmT}\bS_\rmp\bF_\rmp$ in (\ref{receivedsignalatpilot}) is not of full column rank. Thus, the channel estimation in (\ref{LSEestimation}) with $N_\rmp$ pilot sub-carriers may fail. Furthermore, the primary signal detection and thereafter the secondary signal detection can be affected when $\xi  > N_\rmp - L_\rmb - d_\rmb$.

In summary, we consider the case that\footnote{The backscatter link from PTx to CRx via STx experiences double fading with significant attenuation, resulting in some taps having amplitudes that are nearly zeros and can be considered negligible. Thus, we assume that $L_\rmb + d_\rmb \le L_\rmd$. Additionally, the CP length $N_{\rm cp}$ is normally larger than the pilot sub-carrier number $N_\rmp$ \cite{channelestimationofdm}. Besides when $N_{\rm cp} < N_\rmp$, the results can be easily derived accordingly.} $L_\rmb + d_\rmb \le L_\rmd \le N_\rmp \le N_{\rm cp}$ as an example to explain the effect of symbol synchronization error and obtain its tolerable range. When $\xi \in [0, L_\rmd - L_\rmb - d_\rmb]$, there exists no effect on the performance of our proposed system. When $\xi \in (L_\rmd - L_\rmb - d_\rmb, N_\rmp - L_\rmb - d_\rmb]$, the composite-link channel can still be estimated by (\ref{LSEestimation}), (\ref{estimation1}) and (\ref{estimation2}), but the estimation performance will degrade. The primary and secondary signal detection can be affected accordingly. Since our proposed system can still work well when the symbol synchronization error $\xi$ is in the previous two ranges, we obtain the range of tolerable synchronization error as $[0, N_\rmp - L_\rmb - d_\rmb]$. On the other hand, when $\xi \in (N_\rmp - L_\rmb - d_\rmb, N_{\rm cp} - L_\rmb - d_\rmb + 1]$, the channel estimation in (\ref{LSEestimation}) may fail, which has a more significant impact on the primary and secondary signal detection. When $\xi \in (N_{\rm cp} - L_\rmb - d_\rmb + 1, N + N_{\rm cp} - 1)$, besides the previously described impact on the channel estimation in (\ref{LSEestimation}), the introduction of the IBI and ICI can also affect the detection of the primary and secondary signals. In order to avoid $\xi$ falling outside the tolerable range, the CP length $N_{\rm cp}$ and the number of comb-type pilots $N_\rmp$ can be designed to be larger, or STx can be equipped with more accurate synchronization circuits.

\section{Simulation Results}\label{simulation}

\noindent In this section, we present the simulation results to evaluate the performance of our proposed system and validate our BER analysis. 

We consider the primary system which has $N = 64$ OFDM sub-carriers with $N_\rmp = 8$ equally spaced pilot sub-carriers.
The CP length is set as $N_{\rm cp} = 16$. The numbers of taps for the direct link and the backscatter link are set as $L_\rmd = 4$ and $L_\rmb = 2$. The channel taps are modeled as Rayleigh fading with equal average power. 
The large-scale fadings of PTx-CRx, PTx-STx and STx-CRx links are modeled as $\beta_\rmd = 10^{-3}d_{\rmd}^{-v_{\rmd}}$, $\beta_1 = 10^{-3}d_1^{-v_1}$, and $\beta_2 = 10^{-3}d_2^{-v_2}$, respectively. The distance of PTx-CRx link is set as $d_\rmd = 200$ m. STx lies on the line that connects PTx and CRx, resulting in the distances of PTx-STx and STx-CRx links satisfying $d_1 + d_2 = d_\rmd$. The path-loss exponents of PTx-CRx, PTx-STx and STx-CRx links are set as $v_\rmd = 2.5$ and $v_1 = v_2 = 2$, respectively. The noise power is set as $\sigma^2 = -80$ dBm. We define the SNR ratio as $\Delta \gamma = \bar{\gamma}_\rmb / \bar{\gamma}_\rmd = \beta_1\beta_2/\beta_\rmd$, where $\bar{\gamma}_\rmb = P_\rmT\beta_1\beta_2/\sigma^2$ and $\bar{\gamma}_\rmd = P_\rmT\beta_\rmd/\sigma^2$ are the backscatter- and direct-link SNR, respectively. The transmission power $P_\rmT$ is determined by the value of the direct-link SNR $\bar{\gamma}_\rmd$ in the following. Unless otherwise specified, the distance of PTx-STx link $d_1$ is set as $3.83$ m and thus $\Delta \gamma = -30$ dB.
The delay of the backscatter signal is set as $1$ sample. We consider that the primary information signals are modulated by 16-QAM, and the secondary signals are modulated by 8-PSK. The average BER results are obtained over $10^6$ channel realizations.

\begin{figure}[t!]
	\centering
	\includegraphics[width=0.9\columnwidth]{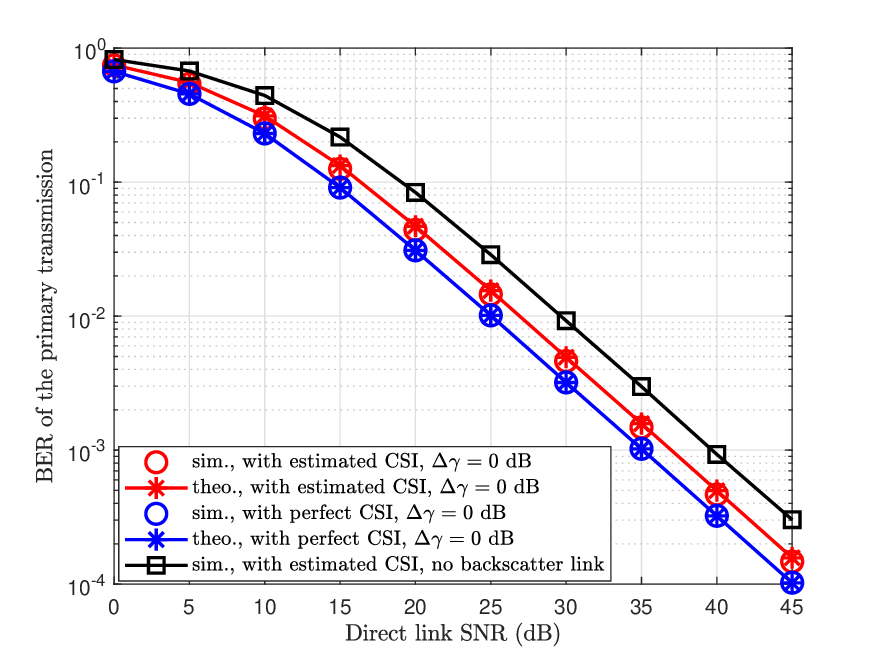}
	\caption{BER of the primary transmission versus the direct link SNR.}
	\label{fig:Fig4}
	\vspace{-0.5cm}
\end{figure}

\begin{figure}[t!]
	\centering 	
	\includegraphics[width=0.9\columnwidth]{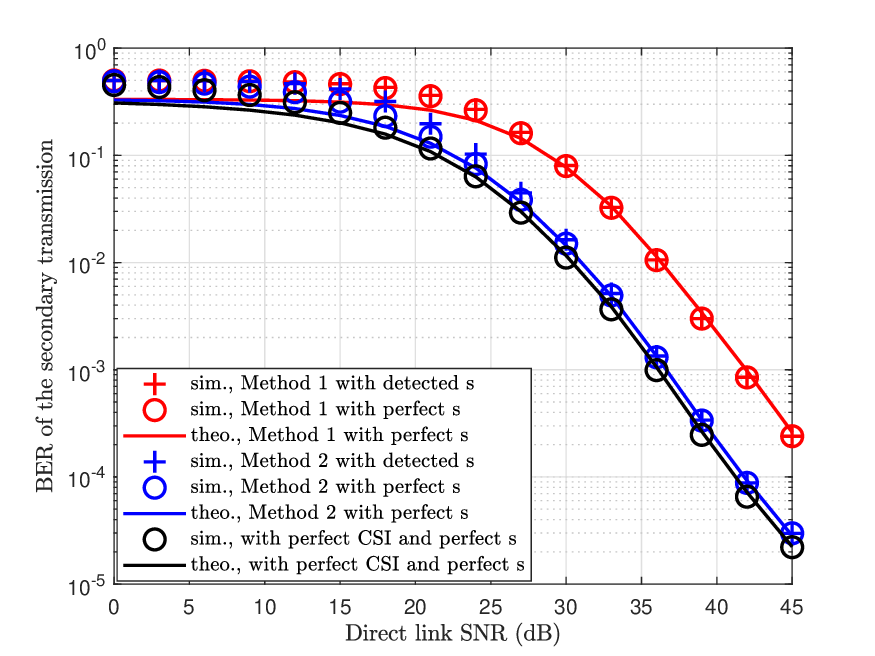}
	\caption{BER of the secondary transmission versus the direct link SNR.}
	\label{fig:Fig5}
	\vspace{-0.5cm}
\end{figure}

Fig.~\ref{fig:Fig4} depicts the BER performance of the primary transmission versus the direct link SNR. The distance from PTx to STx is set as $0.12$ m, and thus the SNR ratio is $\Delta\gamma = 0$ dB. We observe that the theoretical results coincide with the simulated ones for the cases with perfect CSI and estimated CSI. Compared with the case of no backscatter link, we find that the backscatter link can provide a performance gain to the primary transmission. For example, at the BER level of $10^{-3}$, a gain of 3 dB can be achieved when $\Delta \gamma = 0$ dB. On the other hand, the diversity order of the primary transmission is 1 due to the adopted uncoded OFDM scheme.

Fig.~\ref{fig:Fig5} depicts the BER performance of the secondary transmission versus the direct link SNR with our proposed two channel estimation methods.
First, we observe that the theoretical results are consistent with the simulated ones for \emph{Method 1}, as well as the approximate results for \emph{Method 2}. Next, we can see that although the errors of $\hatbS\left(n\right)$ degrade the BER performance of the secondary transmission in the low SNR regime, as the direct link SNR increases, the BERs with the detected primary signal coincide with the BERs with the perfectly detected primary signal. This is because few errors of the primary signal within each OFDM symbol have negligible impact on the detection of $c\left(n\right)$. Moreover, \emph{Method 2} outperforms \emph{Method 1} as analyzed in Section \ref{2c}, since \emph{Method 2} utilizes the channel correlation among all the sub-carriers.

\begin{figure}[t!]
	\centering
	\subfigure[Different SNR ratio $\Delta \gamma$.]{
		\begin{minipage}{0.9\columnwidth}
			\label{fig:Fig7a}
			\includegraphics[width=1\columnwidth]{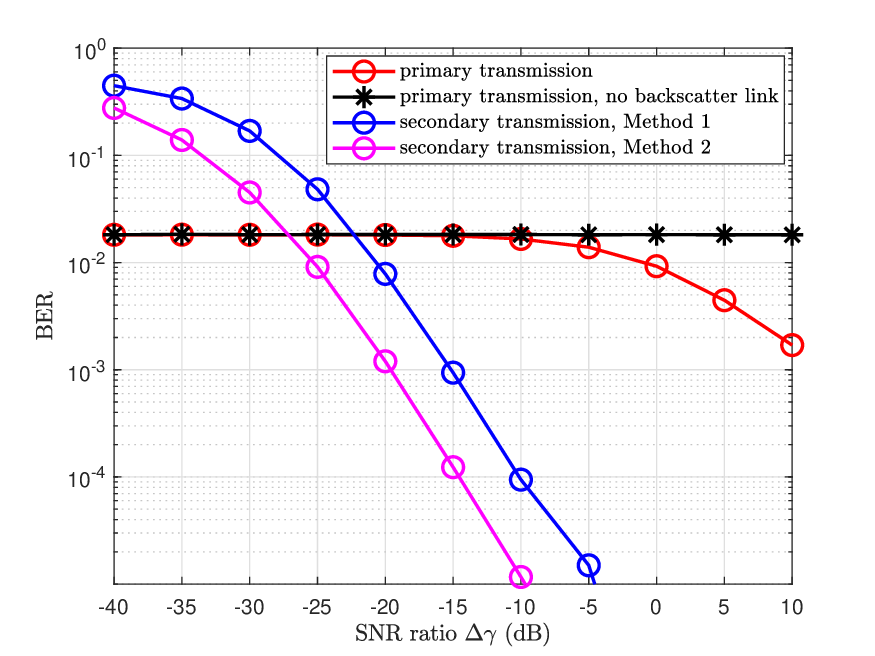}
		\end{minipage}
	}
	\subfigure[Different channel models.]{
		\begin{minipage}{0.9\columnwidth}
			\label{fig:Fig7b}
			\includegraphics[width=1\columnwidth]{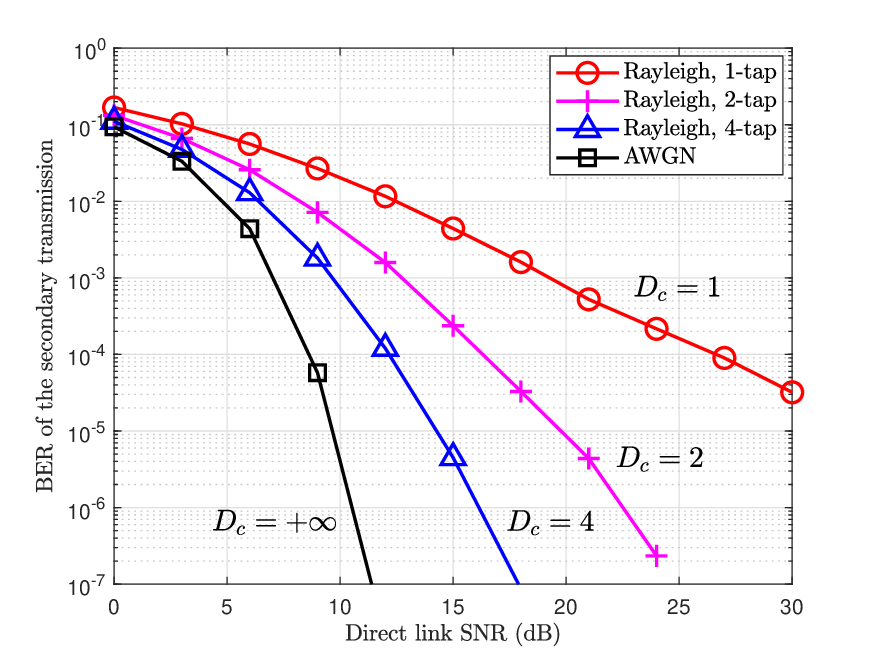}
		\end{minipage}
	}
	\caption{BER of the secondary transmission versus the direct link SNR with different backscatter link channel setups.}
	\label{fig:Fig7}
	\vspace{-0.5cm}
\end{figure}

Fig.~\ref{fig:Fig7a} depicts the BER performance of the primary and secondary transmissions versus the SNR ratio $\Delta\gamma$. The direct link SNR is fixed as $\bar{\gamma}_\rmd = 27$ dB. We observe that better BER performance of the secondary transmission can be achieved with the increase of $\Delta\gamma$ since the stronger backscatter link can lead to a larger SNR of the secondary signal. Furthermore, a BER performance gain to the primary transmission becomes more significant when $\Delta\gamma \ge -10$ dB, which is negligible when $\Delta\gamma < -10$ dB.

Fig.~\ref{fig:Fig7b} depicts the BER performance of the secondary transmission via \emph{Method 2} versus the direct link SNR with different backscatter channel fading settings. Specifically, we consider four settings, which include an AWGN channel and Rayleigh fading channels with 1, 2, and 4 taps. We can see that different channel fading settings can bring different diversity gains to the secondary transmission. In particular, for the 1-, 2-, and 4-tap Rayleigh fading channels, frequency diversity orders of 1, 2, and 4 can be achieved in the high SNR regime with \emph{Method 2}, which is consistent with the analysis in Section \ref{diversityorder}. Combining the observations in Fig.~\ref{fig:Fig7a} and Fig.~\ref{fig:Fig7b}, the performance of the secondary transmission can be improved by enhancing the gain and the number of channel taps for the backscatter link.

\begin{figure}[t!]
	\centering
	\includegraphics[width=0.9\columnwidth]{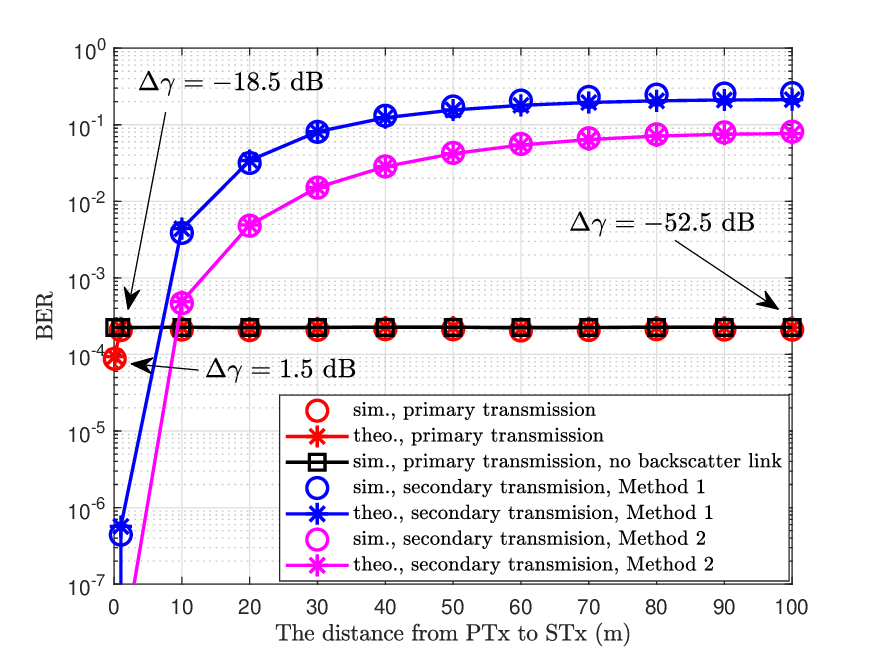}
	\caption{BER of SR versus the distance from PTx to STx.}
	\label{fig:Fig8}
	\vspace{-0.5cm}
\end{figure}

Fig.~\ref{fig:Fig8} depicts the BER performance of the primary and secondary transmissions versus the distance from PTx to STx. Notice that the results of the distance from $100$ m to $200$ m are symmetrical with the results in Fig.~\ref{fig:Fig8}. Firstly, the BER performances of the secondary transmission with the two proposed methods worsen as the distance increases. 
Meanwhile, we note that the SNR ratio also becomes smaller with the increase on the distance.
This is because the multiplicative fading in the backscatter link becomes more severe, and the strength of the backscatter link weakens. Consequently, the backscatter-link SNR decreases, leading to a worse BER performance of the secondary transmission.
Meanwhile, the BER performance gain to the primary transmission turns almost null when the distance increases from $0.1$ m to $1$ m. This indicates that the BER performance can be better for both the primary and secondary transmissions when STx is located close to PTx or CRx.

\begin{figure}[t!]
	\centering
	\subfigure[Primary transmission.]{
		\begin{minipage}{0.9\columnwidth}
			\label{fig:Fig9a}
			\includegraphics[width=1\columnwidth]{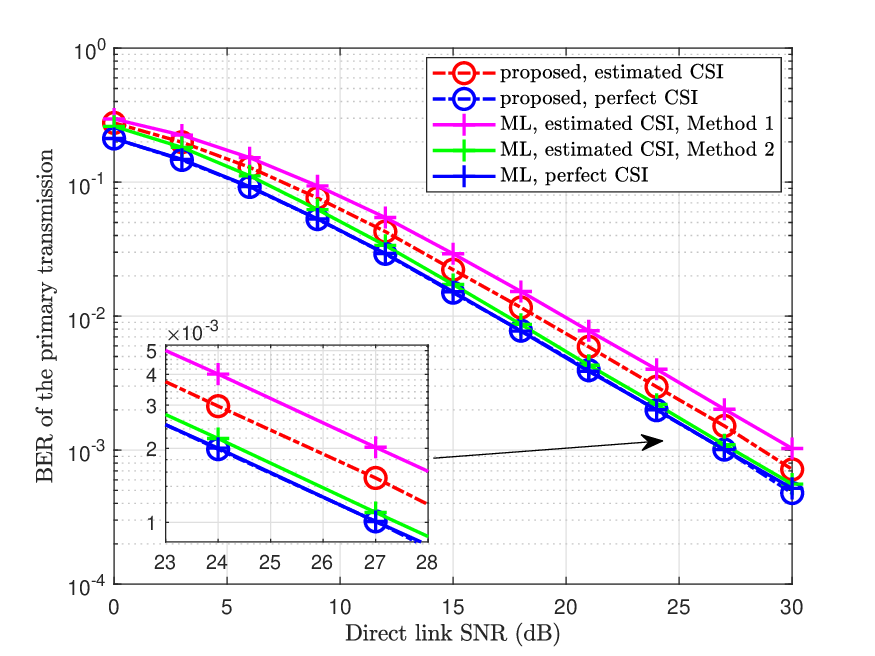}
		\end{minipage}
	}
	\subfigure[Secondary transmission.]{
		\begin{minipage}{0.9\columnwidth}
			\label{fig:Fig9b}
			\includegraphics[width=1\columnwidth]{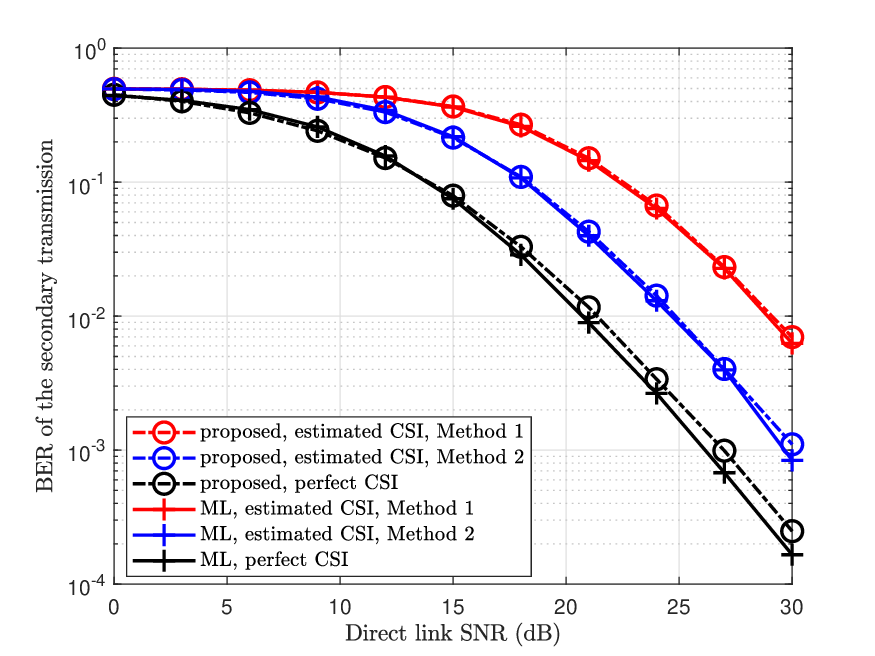}
		\end{minipage}
	}
	\caption{BER of SR versus the direct link SNR by using our proposed receiver and the two-step ML receiver \cite{CABC}.}
	\label{fig:Fig9}
	\vspace{-0.5cm}
\end{figure}

Fig.~\ref{fig:Fig9a} depicts the BER performance of the primary transmission versus the direct link SNR by using our proposed receiver and the two-step ML receiver \cite{CABC}. Since the computation complexity of the ML receiver is high, their modulation orders are set as $M_s = 4$ and $M_c = 2$, respectively. Firstly, we observe that our proposed receiver can achieve almost the same BER performance as the ML receiver when the perfect CSI is acquired. On the other hand, with estimated CSI of \emph{Method 1}, our proposed receiver outperforms the ML receiver. When the accuracy of the estimated CSI is better via \emph{Method 2}, the ML receiver outperforms our proposed receiver slightly. At the BER level of $10^{-3}$, the ML receiver with \emph{Method 2} obtains an SNR gain of about $1$ dB compared with our proposed receiver. 

Fig.~\ref{fig:Fig9b} depicts the BER performance of the secondary transmission versus the direct link SNR by using our proposed receiver and the two-step ML receiver. We notice that our proposed system can achieve almost the same BER performance with the ML receiver when the estimated CSI is used via both \emph{Method 1} and \emph{Method 2}. For the case of perfect CSI, the ML receiver outperforms our proposed receiver slightly in the high SNR regime. In summary, our proposed receiver can achieve the BER performance extremely close to the two-step ML receiver with lower computation complexity.

\begin{figure}[t!]
	\centering
	\includegraphics[width=0.9\columnwidth]{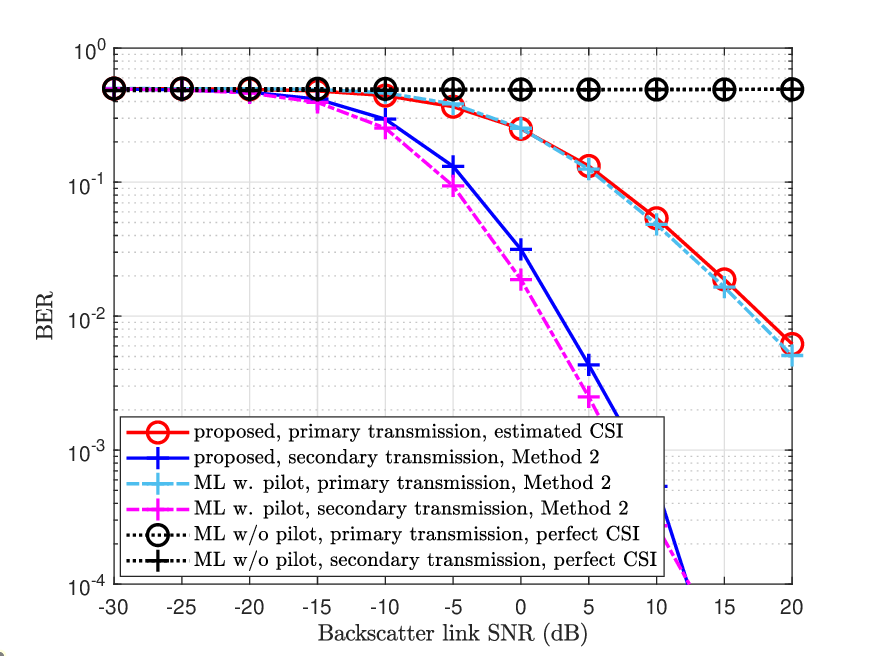}
	\caption{BER of SR versus backscatter-link SNR without direct link.}
	\label{fig:Fig10}
	\vspace{-0.5cm}
\end{figure}

Fig.~\ref{fig:Fig10} depicts the BER performance of the primary and secondary transmissions versus the backscatter link SNR in the case where the direct link is absent. 
We consider the two-step ML receivers with and without our proposed pilot structure as the benchmark schemes.
First, without the use of the pilot structure, the ML receiver cannot detect the primary and secondary signals correctly in the absence of the direct link, as it fails to remove the coupling between the primary and secondary signals in the backscatter link. Conversely, the ML receiver with the pilot structure can successfully detect primary and secondary signals, since our proposed pilot structure aids the ML receiver in decoupling the primary and secondary signals by providing the reference signals within each block of primary signals. Moreover, our proposed receiver is also able to successfully detect the primary and secondary signals thanks to our proposed pilot structure. This is because the secondary signal is treated as a part of the composite channel during the primary signal detection, enabling the primary and secondary signals to be decoupled during detection. 
Additionally, it is observed that the BER of the secondary transmission is much better than that of the primary transmission due to the spreading gain in detecting the secondary signal. Furthermore, the BER curves for the secondary transmission via \emph{Method 1} and \emph{Method 2} are almost the same due to the bad BER performance for the primary transmission in low SNR regime. 

\begin{figure}[t!]
	\centering
	\includegraphics[width=0.9\columnwidth]{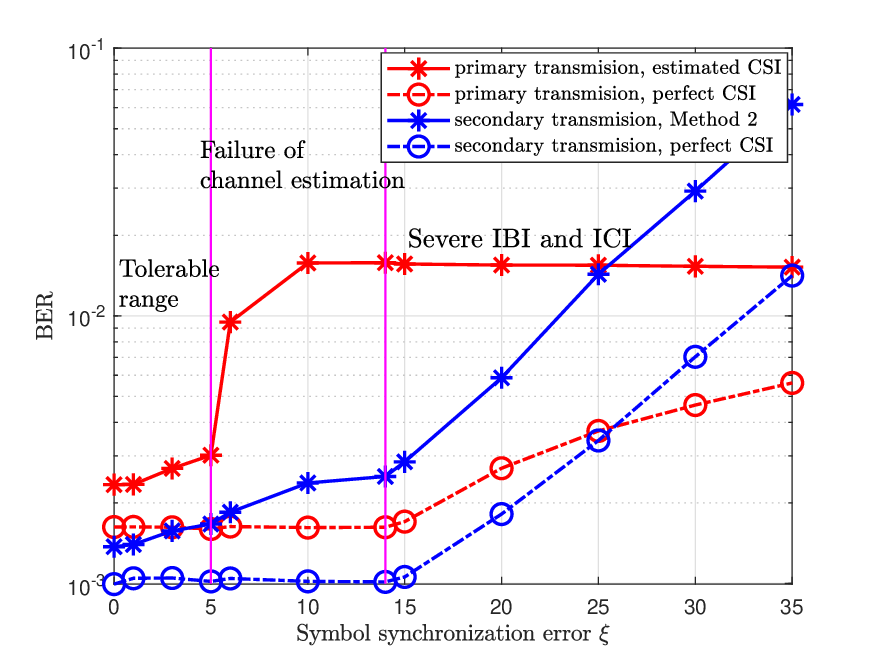}
	\caption{BER of SR versus symbol synchronization error.}
	\label{fig:Fig11}
	\vspace{-0.5cm}
\end{figure}

Fig.~\ref{fig:Fig11} depicts the BER performance of the primary and secondary transmissions via \emph{Method 2} versus the symbol synchronization error $\xi$. Firstly, we observe that the primary and secondary signals can be successfully detected when the error $\xi$ is in the tolerable range. For the case with estimated CSI, the BER performance degrades since the number of taps for composite link $L$ increases and the estimation accuracy decreases. When $5 < \xi \le 14$, the number of comb-type pilots is smaller than $L$, which leads to the failure of channel estimation for primary signal detection. The BER performance of the primary transmission significantly degrades, while the influence on the secondary transmission is negligible. This is because the composite-link channel is estimated again for the secondary transmission with the help of the detected primary signals, the number of which is still larger than $L$. Meanwhile, there exists no effect for the case with perfect CSI. When $\xi > 15$, the insufficient CP results in severe IBI and ICI to the primary and secondary transmissions. The BER performance of both the primary and secondary significantly degrades for the case of both perfect and estimated CSI. To sum up, our analysis of symbol synchronization errors is validated. And our proposed receiver can endure the errors when it falls within the tolerable range.

\section{Conclusions}\label{Conclusions}

\noindent In this paper, we have proposed a pilot design and the signal detection method for an SR system where the primary transmission adopts OFDM. To preserve orthogonality among the OFDM sub-carriers, the secondary signal has been designed to span an entire OFDM symbol. The comb-type pilot design and preamble pilot design have been utilized by the primary and secondary transmissions, respectively. Furthermore, the detection methods of the primary and secondary signals have been studied with pilot-assisted channel estimation. The BER performance with both perfect and estimated CSI, the diversity orders of the primary and secondary transmissions, and the sensitivity to symbol synchronization errors have been analyzed. Simulation results have shown that the proposed system can allow the primary transmission to achieve a BER performance gain while enabling the secondary transmission. In particular, even without the direct link, the primary and secondary transmissions can be supported via only the backscatter link. Additionally, our proposed SR system does not require the secondary signals to have perfect symbol synchronization with the primary ones and it can tolerate some level of synchronization errors.

\begin{appendices}
\section{Proof of Theorem \ref{theorem1}}\label{theorem1proof}

\noindent Stack the estimated composite-link CFRs of the $T$ preamble pilot symbols together as 
\begin{align}\label{linearmodel}
	\underbrace{\begin{bmatrix}
		\hatbH(0) \\
		\vdots\\
		\hatbH(T \!-\! 1)
	\end{bmatrix}}_{\bb} \!\!=\!\! \underbrace{\begin{bmatrix}
	1 & c(0)\\
	\vdots & \vdots\\
	1 & c(T \!-\! 1)
\end{bmatrix} \!\otimes\! \bI_N}_{\bC} \! \begin{bmatrix}
\bH_\rmd \\
\bH_\rmb
\end{bmatrix} \! + \! \begin{bmatrix}
\bepsilon(0) \\
\vdots \\
\bepsilon(T \!-\! 1)
\end{bmatrix}\!\!,
\end{align}
where $\bepsilon(n)$ is the estimation error of $\hatbH(n)$ and we assume that $\bepsilon(n) \sim \calC\calN(\bm{0}, \sigma_\epsilon^2\bI_N)$. Notice that the model in (\ref{linearmodel}) follows the linear model. In order to estimate $[\bH_\rmd^\rmT, \bH_\rmb^\rmT]^\rmT$, the number of preamble pilot symbols needs to satisfy $T\ge2$. The corresponding minimum variance unbiased (MVU) estimator can be derived as the LSE estimator \cite{kay1993fundamentals}. Thus, the CFRs can be estimated as
\begin{align}
	\begin{bmatrix}
		\hatbH_\rmd^\rmT & \hatbH_\rmb^\rmT
	\end{bmatrix}^\rmT = \left(\bC^\rmH\bC\right)^{-1}\bC^\rmH \bb.
\end{align}
Furthermore, the corresponding covariance matrix of the estimation can be derived as 
\begin{equation}
	\begin{split}
		\bSigma &= \sigma_\epsilon^2\left(\bC^\rmH\bC\right)^{-1} \\
		&= \sigma_\epsilon^2\left(\begin{bmatrix}
			T & \sum_{n = 0}^{T - 1} c(n) \\
			\sum_{n = 0}^{T - 1} c^\dagger(n) & \sum_{n = 0}^{T - 1} \left|c(n)\right|^2
		\end{bmatrix} \otimes \bI_N\right) ^ {-1}.
	\end{split}
\end{equation}
To minimize the variance of the estimation or the trace of $\bSigma$, the matrix $\bC^\rmH\bC$ needs to be a diagonal matrix \cite{kay1993fundamentals}, i.e., $\sum_{n = 0}^{T - 1} c(n) = 0$. Furthermore, the minimum variance can be attained with the maximum of the diagonal elements of $\bC^\rmH\bC$, i.e., $|c(n)|^2 = 1$, $n = 0, \dots, T-1$. As a result, the estimation of the direct- and backscatter-link CFRs can be derived as (\ref{estimationwithT}). The covariance matrix can be derived as $\bSigma = \sigma_\epsilon^2\bI_{2N}/T$.

\section{Proof of Theorem \ref{theorem2} and Theorem \ref{theorem3}}\label{theorem23proof}
\noindent We present the proof for \emph{Method 2} here and the proof for \emph{Method 1} can be similarly derived. Recalling the error expression and distribution illustrated in Section \ref{estimationwitherror}, the variance of the zero-mean equivalent noise in (\ref{dss2}) can be derived as
\begin{align} \label{proof2}
	\sigma^2_\rmn \!&= \!\bbE\!\left[\left|\bh_\rmb^\rmH\bepsilon\!\left(n\right)\right|^2\right] \!+\! \bbE\!\left[\left|\bh_\rmb^\rmH\bepsilon_\rmd\right|^2\right] \!+\! \bbE\!\left[\left|\bepsilon_\rmb^\rmH\bepsilon\!\left(n\right)\right|\right] \!+\! \bbE\!\left[\left|\bepsilon_\rmb^\rmH\bepsilon_\rmd\right|\right] \nonumber \\
	& = \!\frac{3\left\|\bh_\rmb\right\|^2\sigma^2}{2P_\rmT} + \bbE\left[\left|\bepsilon_\rmb^\rmH\bepsilon\left(n\right)\right|\right] + \bbE\left[\left|\bepsilon_\rmb^\rmH\bepsilon_\rmd\right|\right].
\end{align}
Recall that $\bepsilon_\rmd = (\bepsilon(0) + \bepsilon(1))/2$ and $\bepsilon_\rmd = (\bepsilon(0) + \bepsilon(1))/2$. Since $\bepsilon\left(n\right)$ is i.i.d. with respect to $n$, the last two terms in (\ref{proof2}) can be derived as
\begin{multline}\label{proof3}
	\bbE\left[\left|\bepsilon_\rmb^\rmH\bepsilon\left(n\right)\right|\right] + \bbE\left[\left|\bepsilon_\rmb^\rmH\bepsilon_\rmd\right|\right] = \frac{5}{8} \bbE\left[\left|\bepsilon^\rmH\left(0\right)\bepsilon\left(n\right)\right|^2\right] + \\
	\frac{1}{8} \bbE\left[\left|\bepsilon^\rmH\left(0\right) \bepsilon\left(0\right)\right|^2\right] - \frac{1}{8}\left(\bbE\left[\bepsilon^\rmH\left(0\right)\bepsilon\left(0\right)\right]\right)^2.
\end{multline}
We take the term $\bbE[|\bepsilon^\rmH(0)\bepsilon(0)|^2]$ in (\ref{proof3}) for example. With \emph{Method 2}, the estimation error can be expressed by 
\begin{align}
	\bepsilon(0) = \bF_L \bG_2 \bU(0) = \frac{1}{N\sqrt{P_\rmT}}\bF_L\bF_L^\rmH\bS^\rmH(0)\bU(0).
\end{align}
Recall the fact that $\bF_L^\rmH\bF_L = N\bI_L$. By replacing the expression of $\bepsilon\left(0\right)$, we can obtain
\begin{align}
	&\bbE\left[\left|\bepsilon^\rmH\left(0\right)\bepsilon\left(0\right)\right|^2\right] = \bbE\left[\bepsilon^\rmH(0)\bepsilon(0)\bepsilon^\rmH(0)\bepsilon(0)\right] \nonumber\\
	= & \frac{1}{N^2P_\rmT^2}\bbE\left[\left(\bU^\rmH(0)\bS(0)\bF_L\bF_L^\rmH\bS^\rmH(0)\bU(0)\right)^2\right] \nonumber \\
	= & \frac{1}{N^2P_\rmT^2}\bbE \left[\left(\sum_{n = 0}^{N - 1}\sum_{k = 0}^{N - 1}U_k^\dagger\left(0\right) S_k\left(0\right)\boldf_k^\rmH\boldf_n U_n\left(0\right) S_n^\dagger\left(0\right)\right)^2\right]\nonumber \\
	\overset{(a)}{=} & \frac{1}{N^2P_\rmT^2}\left(\sum_{k_1 = 0}^{N - 1}\sum_{k_2 \neq k_1}^{N - 1}\bbE\left[\left|U_{k_1}\left(0\right)\right|^2\left|U_{k_2}\left(0\right)\right|^2\right]\boldf_{k_1}^\rmH\boldf_{k_1}\boldf_{k_2}^\rmH\boldf_{k_2} + \right. \nonumber \\
	& \left. \sum_{k_1 = 0}^{N - 1}\sum_{k_2 \neq k_1}^{N - 1}\bbE\left[\left|U_{k_1}\left(0\right)\right|^2\left|U_{k_2}\left(0\right)\right|^2\right]\boldf_{k_1}^\rmH\boldf_{k_2}\boldf_{k_2}^\rmH\boldf_{k_1} + \right. \\
	& \left. \sum_{k = 0}^{N - 1}\bbE\left[\left|U_k\left(0\right)\right|^4\right]\boldf_k^\rmH\boldf_k\boldf_k^\rmH\boldf_k\right) \nonumber \\
	\overset{(b)}{=} & \frac{\sigma^4}{N^2 P_\rmT^2}\left( N \left(N \!-\! 1\right) L^2 \!+\! \left(\left\|\bF_L\bF_L^\rmH\right\|_\rmF^2 \!-\! N L^2\right) \!+\! 2 N L^2  \right) \nonumber\\
	\overset{(c)}{=} & \frac{\left(L^2 + L \right)\sigma^4 }{P_\rmT^2},
\end{align}
where $(a)$ holds from the fact that $U_k\!\left(0\right)$ is i.i.d. with respect to $k$, $(b)$ holds from $\boldf_k^\rmH\boldf_k = L$ and $\bbE[|U_k(0)|^4] = 2\sigma^4$, and $(c)$ holds from $\|\bF_L\bF_L^\rmH\|_\rmF^2 = N^2L$. Similarly, we can obtain that $\bbE[|\bepsilon^\rmH(0)\bepsilon(n)|^2] = L\sigma^4/P_\rmT^2$ and $\bbE[\bepsilon^\rmH(0)\bepsilon(0)] = L\sigma^2/P_\rmT$. Thus, $\sigma_\rmn^2$ can be derived, as well as the SNR expression of the case with \emph{Method 2}.

\end{appendices}

\bibliography{IEEEabrv,reference}

\bibliographystyle{IEEEtran}

\end{document}